\documentclass[twocolumn,prc,floatfix,showkeys,showpacs,preprintnumbers,amsmath,amssymb]{revtex4}

\usepackage{graphicx}
\usepackage{dcolumn}
\usepackage{bm}
\usepackage{multirow} 

\begin{document}

\title{Microscopic description of cluster radioactivity in actinide nuclei}

\author{M. Warda}
\affiliation{Katedra Fizyki Teoretycznej, Uniwersytet Marii Curie--Sk\l odowskiej,
        ul. Radziszewskiego 10, 20-031 Lublin, Poland}
\email{warda@kft.umcs.lublin.pl}

\author{L.M. Robledo}
\affiliation{Departamento de F\'\i sica Te\'orica (M\'odulo 15), 
Universidad Aut\'onoma de Madrid, E-28049 Madrid, Spain}
\email{luis.robledo@uam.es}

\date{\today}

\begin{abstract}
Cluster radioactivity is the emission of a fragment heavier than $\alpha$
particle and lighter than mass 50. The range of clusters observed in experiments goes from  
$^{14}$C to $^{32}$Si while the heavy mass residue is always a 
nucleus in the neighborhood of the doubly-magic $^{208}$Pb nucleus.
Cluster radioactivity is described in this paper as a very 
asymmetric nuclear fission. 
A new fission valley leading to a decay with large fragment 
mass asymmetry matching the cluster radioactivity products is found.
The mass octupole moment is 
found to be more convenient than the standard quadrupole moment as 
the parameter driving the system to fission. The mean-field HFB theory with 
the phenomenological Gogny interaction has been used to compute the 
cluster emission properties of a wide range of even-even actinide nuclei from 
$^{222}$Ra to $^{242}$Cm, where emission of the clusters has been 
experimentally observed.  
Computed half-lives for cluster emission are compared with 
experimental results. The noticeable agreement obtained between the  
predicted properties of cluster emission (namely, clusters masses and emission half-lives) and the measured data confirms the 
validity of the proposed methodology in the analysis of the 
phenomenon of cluster radioactivity. 
A continuous fission path through the scission point has been described using the neck parameter constraint.
\end{abstract}

\keywords{
cluster radioactivity, actinide nuclei, nuclear fission, octupole moment}

\pacs{
23.90.+w, 	
25.85.Ca, 	
27.90.+b 	
}
\maketitle

\section{Introduction}

The emission of $\alpha$ particles and nuclear fission are the two 
dominant spontaneous decay modes of heavy and super-heavy nuclei. In 
both cases two nuclei are produced. 
In  $\alpha$ decay one  $^{4}$He nucleus ($\alpha$ particle) 
is emitted out of the parent nucleus and the remaining nucleons form 
a heavy mass residue with $N-2$ neutrons and $Z-2$ protons. In 
contrast to the huge mass asymmetry of $\alpha$ decay, two nuclei of 
similar mass are created in nuclear fission. A large  variety of 
isotopes are produced in spontaneous fission with masses covering 
the range from $A\sim70$ to  $A\sim190$. In many heavy nuclei the 
dominant decay channel corresponds to asymmetric fission with the 
most probable  mass of heavy fragment $A\sim140$ and the mass of the 
light one in the range from $A\sim100$ to $A\sim120$ depending on 
the mass of parent nucleus. Symmetric fission is also possible 
in some nuclei with the most probable division into two similar 
fragments. Daughter nuclei lighter than $A\sim60$ have never been 
observed in any fission experiment. Therefore there is a clear 
distinction between  $\alpha$ emission and fission regarding the 
mass of the lighter products as it shows a gap of light nuclei with 
$A\sim 10-50$ that can not be produced in none of the two mentioned 
decay channels of any heavy nuclei. The source of the observed 
differences can be easily explained from basic nuclear properties, 
namely the energetic balance of the two reactions. Fission is 
favorable energetically because of the linear decreasing behavior of 
the binding energy per nucleon for mass numbers larger than  $A\sim 
60$ (the iron peak) that prevents fragments with mass numbers lower 
than that value. On the other hand, $\alpha$ decays energetics is 
dominated by the huge binding energy (as compared to neighboring 
nuclei) of the $\alpha$ particle. 

A common aspect of fission and $\alpha$ decay is that the 
dynamical evolution from the parent nucleus to the daughter
is not favorable energetically, although the $Q$ value of both reactions is positive. Therefore the quantum 
mechanics mechanism of tunneling through a potential barrier is required
to explain both types of decay. As tunneling probabilities depend
exponentially on the width and height of the barrier the expected 
half-lives can span a wide range of many orders of magnitude. This
peculiarity makes the understanding of fission and $\alpha$ decay 
very challenging.

In 1984 Rose and Jones \cite{ros84} observed for the first time
the emission of the $^{14}$C nucleus from the $^{223}$Ra probe. This 
discovery represented a milestone in the description of nuclear 
radioactivity as it bridged the gap between the $\alpha$ emission 
radioactivity and the standard fission reaction. Since then, 
cluster radioactivity (CR) has been found in twelve even-even isotopes \cite{pri85,bar85,hou85,bar86,wan87,wan89,moo89,tre89,ogl90,wes90,bon90,hus91,bon91,hou91,bon93,tre94,hus95,bon95,ogl00,bon01}
and seven odd-even isotopes (see e.g. references in Ref. \cite{bon99,poe02}) in the  
actinide region.
They range from $^{221}$Fr up to $^{242}$Cm. 
The emission of $^{14}$C, $^{20}$O, $^{23}$F, $^{24-26}$Ne, 
$^{28-30}$Mg and $^{32,34}$Si has been observed. The common 
factor of all cluster radioactivity events is the heavy-mass residue 
which is in the neighborhood of the doubly-magic $^{208}$Pb. 
This fact allows us to better characterize CR as ``lead radioactivity" 
and indicates strong influence of shell effects on the nature of this phenomenon. 

Experiments aiming to find CR in the distant region of the neutron
deficient Ba isotopes have been described in Ref. \cite{oga94,gug95,gug96}. 
In this case another doubly-magic nucleus, namely $^{100}$Sn, 
can be considered as the heavy residue and the carbon isotopes around 
$^{12}$C are expected to be emitted. The experiments did not provide 
 evidence for CR in this region  and quantitatively they only 
gave lower limits for the branching ratios for $^{12}$C emission.

CR is an exotic process. The partial 
half-lives are very long and vary in the wide range from $10^{11}$~s to 
$10^{26}$~s. Branching ratios to the dominant $\alpha$ decay in these nuclei 
are very small and are comprised between $10^{-9}$ 
to $10^{-16}$. Moreover spontaneous 
fission is also a competing decay channel in some heavy cluster emitters \cite{aud03a}. 
These reasons clearly 
justify why CR was experimentally discovered as late 
as 45 years after the first fission events which were reported back in 1939 \cite{hah39}.
In the last few decades and thanks to both the interest raised by the 
phenomenon and the impressive improvement of experimental techniques many
examples of CR have been found in several actinide nuclei.
Various experimental methods have been applied to detect the products of 
cluster emission \cite{bon99,tre03,poe02}. First observations were based 
on techniques borrowed from the $\alpha$ decay studies. A $\Delta E-E$ 
telescope made of silicon detectors was used by Rose and Jones in the 
first experiment \cite{ros84}. This method was inconvenient due to 
huge $\alpha$ radioactivity background which could even destroy the  
experimental set-up. Later,  a magnetic field was applied to remove 
the background of charged $\alpha$ particles. Another method used 
in experiments was the detection of 
gamma rays emitted from  exited clusters. Numerous clusters were 
identified in the solid state nuclear track detectors. In this technique 
plastic or glass layers absorb the ionized cluster emitted from the 
radioactive probe. The material of the layer can not be sensitive to 
$\alpha$ radiation and plastic or glass materials with 
proper ionization thresholds are the standard choices.  After irradiation,
the layer is etched to enlarge the track created by the emitted cluster 
as to be visible and well defined under the microscope. The analysis of 
the geometry of the track allows to  identify the emitted cluster.

In the theoretical side, the first successful theoretical 
description of cluster decay was made by Sandulescu {\it et al.}  
\cite{san80} four years before the experimental discovery of this 
reaction. Since the pioneer work of Sandulescu, numerous theoretical 
papers devoted to this end have been published \cite
{egi04,rob08a,rob08b,war11,buc90,ble91,poe91,gup94,kum97,roy98,buc99,maz00,roy01,bas02,
bal04,hor04,iwa04,ogl04,tav07,ni08,qi09,rou09,poe10,san10,bir10,poe11,she11,shi11}. A 
thorough overview of most of the theoretical (mostly 
semi-microscopic) methods can be found in Refs. \cite{poe02,poe96,hoo05,bec10}.
 
As CR is a decay mode ``in between"  $\alpha$ emission and nuclear 
fission, methods already known to both of them can be used to 
describe cluster radioactivity. For instance, the Gamow model of 
$\alpha$ emission  can be extrapolated to describe the emission of 
heavier clusters. It requires the assumption  that a cluster of 
nucleons is preformed in the interior of parent nucleus and then it 
tunnels the barrier of nuclear and Coulomb potential. In this way a 
kind of universal decay law similar to the Geiger-Nuttall formula 
for $\alpha$ emission can be formulated \cite{poe02,qi09,poe11}. The 
main drawback of this approach is that the preformation of the 
cluster inside the parent nucleus is a poorly known and hard to 
characterize process. Nevertheless the half-lives predicted by this 
method agree very well with the experimental data. 

The other method treats cluster radioactivity as a very asymmetric 
fission reaction (see e.g. Refs. \cite{san80,poe96,bec10}). 
The formation of the cluster is a direct consequence of a specific 
kind  of deformation of the parent nucleus. In this approach a 
fission barrier with a specific mass division must be determined. 
Locally maximal barrier transition probability for the specific 
fragments with large mass asymmetry points out for possibility of 
fission with cluster as one of the fragments. Usually the potential 
energy surface (PES) has to be determined as a function of the 
relevant deformation parameters, including elongation and reflection 
asymmetry coordinates. The path in this multidimensional deformation 
surface leading to fission with large fragment mass asymmetry has to 
be found and, finally, the fission barrier must be specified.

We want to show that CR can be fully described microscopically as a 
very asymmetric fission process. We apply standard methods used in 
the theoretical description of nuclear fission which are well 
established in the literature \cite 
{war02,war05a,war05b,war09,ber84,ber90,ber91,del06,dub08}. We use 
the mean-field approximation in the Hartree-Fock-Bogoliubov (HFB) 
scheme with the finite range Gogny force \cite{dec80} to compute the 
nuclear wave functions. Axial symmetry of the nuclear system is 
assumed all along the calculations. Constrains on the quadrupole and 
octupole moments allows to control simultaneously the elongation and 
reflection asymmetry of the system as it evolves to the scission 
point. The wave functions obtained in this way can be used to calculate 
the necessary quantities (energies, collective masses, etc) for a 
physical description of the process. Moreover extra constraint on 
the number of nucleon in the neck (the neck thickness) has been used 
to control the  density distribution around a scission point. 
Description of CR is possible thanks to the identification of a new 
valley in the PES leading to hyper-asymmetric fission. Charge and 
mass numbers of the light fragment created at the hyper-asymmetric 
scission point correspond to what is observed experimentally for a 
given nucleus. Contrary to the standard fission path where the 
leading coordinate is the quadrupole moment, in the hyper-asymmetric 
fission path the relevant coordinate turns out to be the octupole 
moment. Therefore, in our description of CR, all physical quantities 
will be given in  terms of the octupole moment.

First results obtained in this approach have been published in the 
previous papers \cite{egi04,rob08a,rob08b,war11}. CR of selected 
nuclei have been discussed there with some approximations of the 
model.

%
%

In this paper we want to investigate from a microscopic perspective 
all even-even actinide nuclei where cluster radioactivity has been 
experimentally detected. There are twelve such isotopes, namely: 
$^{222,224,226}$Ra, $^{228,230}$Th, $^{230,232,234,236}$U, 
$^{236,238}$Pu, and $^{242}$Cm. Moreover, other three nuclei  
($^{226,232}$Th,  $^{240}$Pu), where experiments have only provided 
lower limits for half-lives of CR, have been examined.

The structure of the paper is as follows: in  Sec. \ref{sec2} the  
theoretical model used in this investigation is described in details. 
Two typical and representative examples of cluster radioactivity 
corresponding to the parent nuclei  $^{224}$Ra and $^{238}$Pu are thoroughly
discussed in Sec. \ref{sec3} as to establish the relevant physics driving the cluster
emission process. Results for 
all the fifteen nuclei considered in this paper are presented in Sec. \ref{sec4}. 
We conclude in Sec. \ref{sec5} with the main consequences extracted from our 
theoretical description.

\section{Theory}
\label{sec2}

As a first step in our theoretical description of cluster emission 
we solve the mean field HFB equation \cite{rin80} with 
the usual constraints on the average number of particles and, in the 
present case, with a constraint on the value of the mass multipole 
moments $\langle Q_{n0}\rangle=Q_{n}$ to analyze the physical 
contents of the process. The axial quadrupole ($Q_2$), octupole
($Q_3$) and hexadecapole ($Q_4$) moments are defined through
the standard Legendre polynomials
\begin{equation}
\hat Q_\lambda=r^\lambda P_\lambda (\cos (\theta))\,.
\end{equation}

The non-linear HFB equation is solved using 
the gradient method \cite{egi95} and taking into account 
approximately second order curvature effects \cite{war02,rob11}. 
The HFB quasiparticle creation and 
annihilation operators are expanded in a harmonic oscillator (HO) 
basis and special attention is paid to the convergence of the 
results with the basis size (see Appendix \ref{appa} for further details). The 
interaction used is the finite range Gogny force with the D1S 
parameterization \cite{ber84}. This interaction has proven to 
successfully describe the fission process in heavy nuclei 
\cite{ber89,ber90,egi00,ber00,war02,del06,dub08,mar09,per09,you09}. 
The other Gogny forces, developed recently: 
D1N  \cite{cha08} and D1M \cite{gor09} are discussed in 
Appendix \ref{appb}. Other details of the HFB calculations are as 
follows: the two body kinetic energy correction (2bKEC) has been included in 
the minimization process. The exchange Coulomb 
contribution is evaluated in the Slater approximation.

All calculations have been performed in the axially symmetric 
regime. It seems to be a rational choice as the systems studied tend 
to be built from a large spherical part reproducing properties of 
doubly-magic nuclei with a small additional part. The lighter 
fragment is often spherical in the ground state. Therefore, the 
influence of non--axial effects is expected to be rather small, if 
any, and may only affect the shape of the barrier just before 
scission reducing slightly its height.


To evaluate the PES we take into account 
correlation energies beyond the mean field. To this end we subtract 
from the HFB energy the rotational energy corrections (REC) stemming 
from the restoration of the rotational symmetry. This correction has 
a considerable influence on the energy landscape (and therefore on 
the height of fission barriers) as it is proportional to the degree 
of rotational symmetry breaking. A full calculation of the REC would 
imply the evaluation of the  angular momentum projected energy \cite
{rod00,rod02}. Unfortunately this kind of beyond mean field 
calculations is only feasible for light nuclei with present day 
computer capabilities. In order to estimate the REC we have followed 
the usual recipe \cite{rin80} (which is well justified for strongly 
deformed configurations) of subtracting to the HFB energy the 
quantity $\langle\Delta\vec{J}^{2}\rangle/(2\mathcal{J}_{Y})$, where 
$\langle\Delta\vec{J}^{2}\rangle$ is the fluctuation on angular 
momentum of the HFB wave function and $\mathcal{J}_{Y}$ is the 
Yoccoz moment of inertia \cite{egi03}. This moment of inertia has 
been computed using the {}``cranking'' approximation in which the 
full linear response matrix appearing in its expression is replaced 
by the zero order approximation (that is, the sum of two 
quasiparticle energies). The impact of this approximation in the 
value of the Yoccoz moment of inertia was analyzed with the Gogny 
interaction for heavy nuclei in \cite{egi00} by comparing the 
approximate value with the one extracted from a complete angular 
momentum projected calculation (see also \cite{rod00} for a 
comparison in light nuclei). The conclusion is that, for strongly 
deformed configurations, the exact REC is roughly a factor 0.7 
smaller than the one computed with the {}``cranking" approximation 
to the Yoccoz moment of inertia. It has also to be mentioned that a 
similar behavior has been observed for the differences between the 
Thouless-Valatin moment of inertia computed exactly and in the 
{}``cranking'' approximation \cite{Gia80,Gir92}. We have taken this 
phenomenological factor into account in our calculation of the REC.

In Sec. \ref{sec4} we will discuss half-lives corresponding to the 
cluster emission and compare them with experimental data. The 
half-lives for cluster emission are computed (in seconds) using the 
standard WKB framework  \cite{hil53}

\begin{equation}
t_{1/2}=2.86\,10^{-21}(1+\exp(2S)).
\end{equation}
The quantity $S$ entering this expression is the action along the 
$Q_{3}$ constrained path 

\begin{equation}
S=\int_{a}^{b}dq_{3}\sqrt{2B(Q_{3})(V(Q_{3})-E_{0})}.
\end{equation}
For the collective inertia $B(Q_{3})$ we have used the ATDHFB 
expression computed again in the {}``cranking" approximation and 
given by \cite{Gia80} 

\begin{equation}
\label{batdhfb}
B_{\mathrm{{ATDHFB}}}(Q_{3})=\frac{M_{-3}(Q_{3})}{M_{-1}^{2}(Q_{3})}
\end{equation}
with the moments $M_{-n}$ given by 

\begin{equation}
M_{-n}(Q_{3})=\sum_{\mu\nu}\frac{\left|\left(Q_{30}^{20}\right)_{\mu\nu}\right|^{2}}
                                        {(E_{\mu}+E_{\nu})^{n}}
\end{equation}
In this expression, $\left(Q_{30}^{20}\right)_{\mu\nu}$ is the 2-quasipartice 0-hole 
component of the octupole operator $\hat{Q}_{30}$ in the 
quasiparticle representation \cite{rin80} and $E_{\mu}$ are the one 
quasiparticle excitation energies obtained as the eigenvalues of the 
HFB hamiltonian matrix.

In the expression for the action 
$V(Q_{3})=E_{HFB}(Q_{3})-REC(Q_{3})-\epsilon_{0}(Q_{3})$ is given by 
the HFB energy minus the REC and the zero point energy (ZPE) 
correction $\epsilon_{0}(Q_{3})$ associated with the octupole 
motion. This ZPE correction is given by
 
\begin{equation}
\epsilon_{0}(Q_{3})=\frac{1}{2}G(Q_{3})B_{\mathrm{{ATDHFB}}}^{-1}(Q_{3})
\end{equation}
where 

\begin{equation}
G(Q_{3})=\frac{M_{-2}(Q_{3})}{2M_{-1}^{2}(Q_{3})}
\end{equation}

Finally, in the expression for the action an additional parameter 
$E_{0}$ is introduced. This parameter can be taken as the HFB energy 
of the (metastable) ground state. However, it is argued that in a 
quantal treatment of the problem the ground state energy is given by 
the HFB energy plus the ZPE associated to the 
collective motion. To account for this fact, the usual recipe is to 
add an estimation of the ZPE to the HFB energy in 
order to obtain $E_{0}$. In our calculations we have considered
this ZPE as a phenomenological parameter
and given a reasonable value of 0.5 MeV for all the isotopes considered \cite{pom08}.

\section{Cluster radioactivity in $^{224}$Ra and $^{238}$Pu}
\label{sec3}

The analysis of CR requires the determination of 
the PES for each nuclei considered in this article. After performing 
these calculations we have found that there are no substantial 
qualitative differences between the various actinide isotopes 
considered. In all cases the PES is similar and only 
quantitative variations are found. Therefore we will not describe in 
details the PES of all actinides. In this section we will 
concentrate only on the CR of two representative 
nuclei, namely the light cluster emitter $^{224}$Ra in which emission of 
$^{14}$C is observed \cite{pri85,hou91} and one of the heaviest emitters
$^{238}$Pu which decay producing the  relatively large clusters 
$^{28,30}$Mg and $^{32}$Si \cite{wan89}. A detailed account of our
previous calculations in some other isotopes can be found in 
\cite{egi04,rob08a,rob08b,war11}.

\subsection{The PES and shapes of fissioning nuclei}
\label{sec3a}
\begin{figure}

\includegraphics[width=0.75\columnwidth, angle=270]{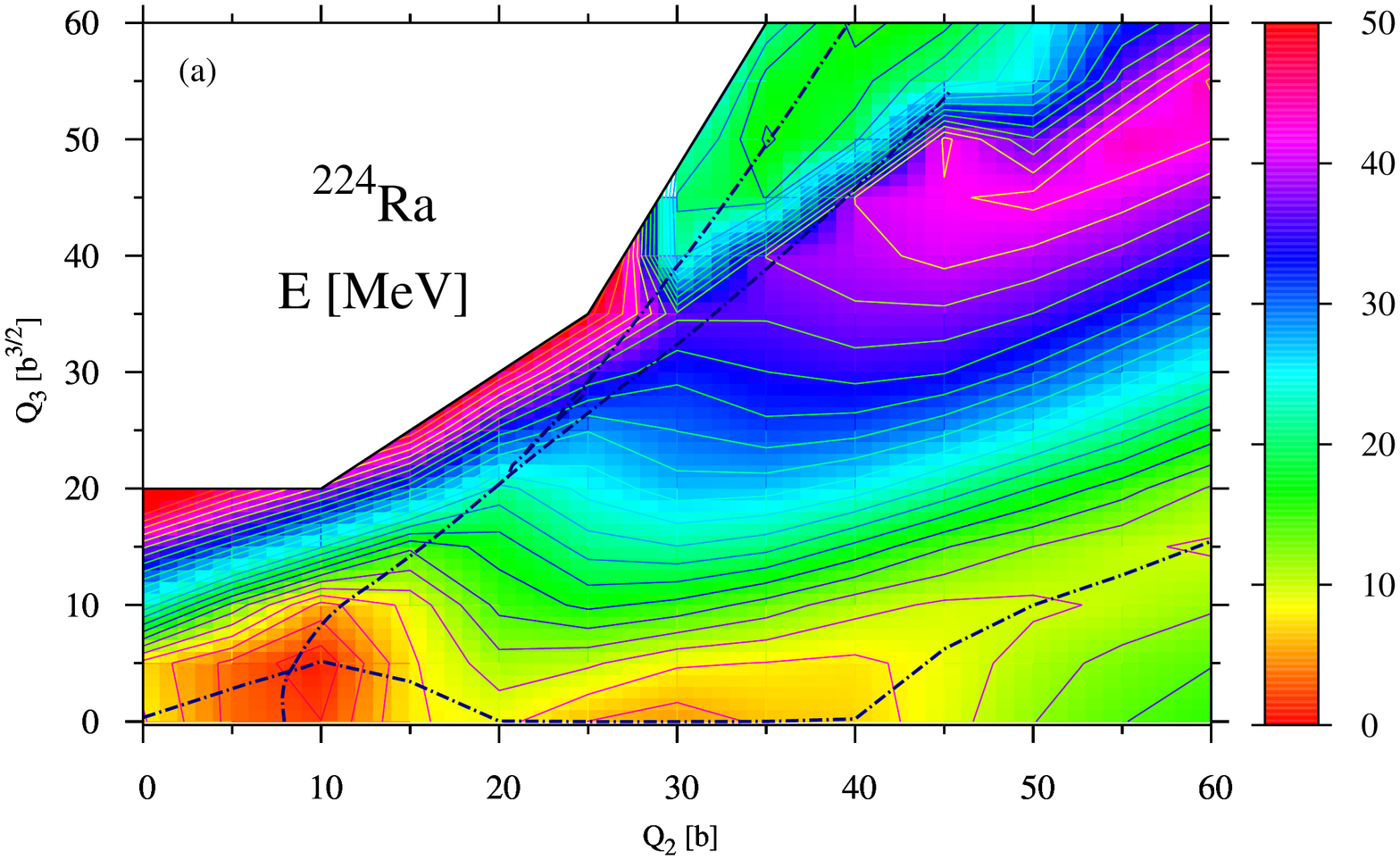}

\includegraphics[width=0.75\columnwidth, angle=270]{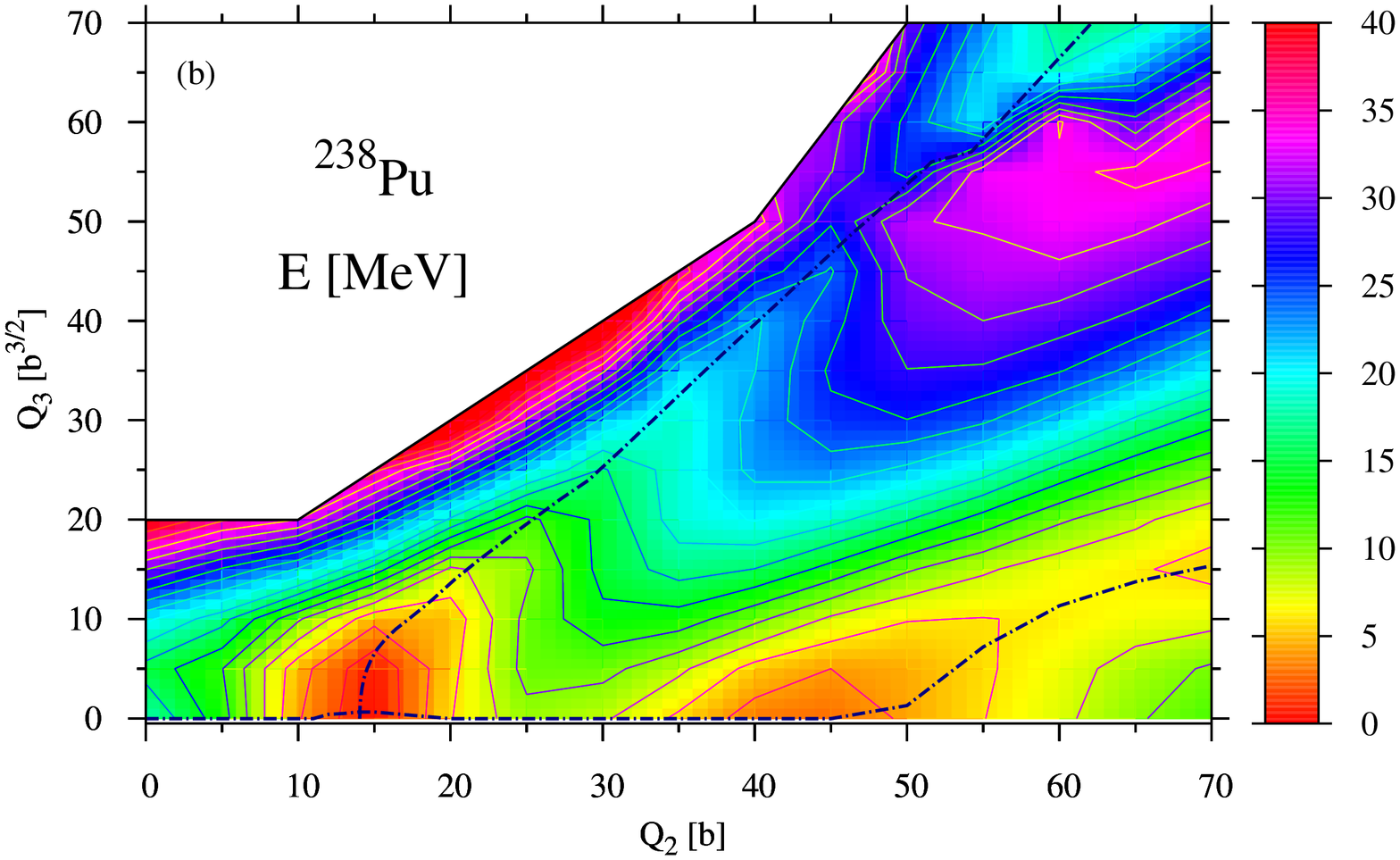}
\caption{\label{map} (Color online) The maps of the PES of (a) $^{224}$Ra and (b)  
$^{238}$Pu as a function of quadrupole moment $Q_2$ and octupole 
moment $Q_3$. Lines of constant energy are plotted every 2 MeV. 
Bold dot-dashed lines are plotted along fission paths.}
\end{figure}

\begin{figure*}
\includegraphics[width=1.0\columnwidth,angle=270]{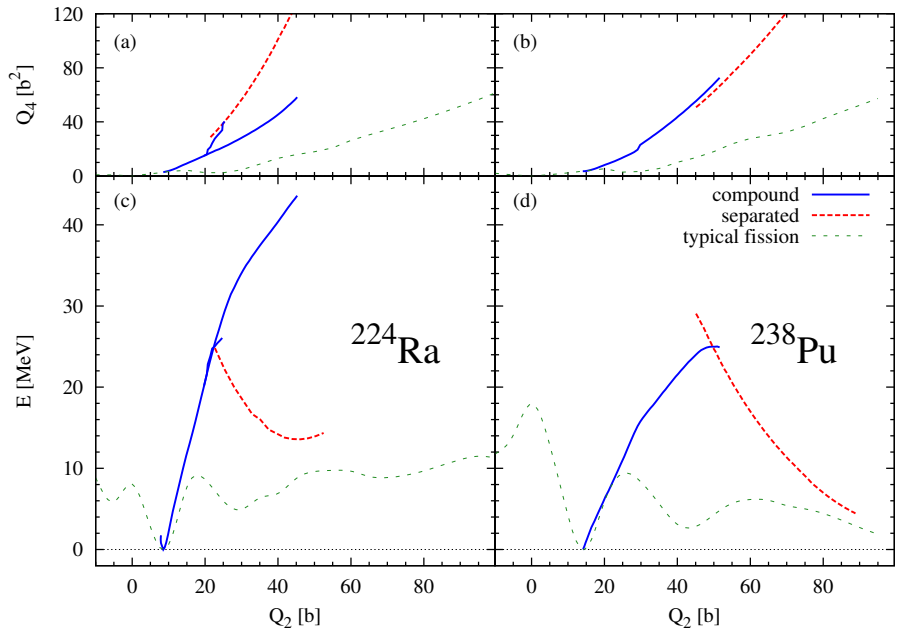}
\caption{\label{barq2} (Color online) Fission barriers in $^{224}$Ra (left) 
and $^{238}$Pu (right) are plotted as a function of the quadrupole moment 
$Q_2$ (lower panels). The values of the hexadecapole moment $Q_4$ of the 
nuclei  along the fission paths are plotted as a function of quadrupole moment $Q_2$  
in the upper panels.}
\end{figure*}

In Fig. \ref{map} we show the PES of $^{224}$Ra and $^{238}$Pu in 
the deformation space of the quadrupole $Q_2$ and octupole $Q_3$ 
moments. This Figure shows how the energy of the system changes with 
the simultaneous changes of elongation (controlled by $Q_2$) and reflection asymmetry (governed by  $Q_3$). Calculations 
have been performed on a grid, with a spacing of 5 b in the $Q_2$ 
direction and of 5 b$^{3/2}$ in the $Q_3$ direction. The oscillator 
lengths characterizing the single particle basis have been optimized 
in every mesh point to minimize the total HFB energy. All the 
values of potential energies  presented in this paper are the 
corresponding HFB energies corrected by the correlation energies of 
the two body kinetic energy correction (2bKEC) and the rotational 
energy correction (REC) as it was described in Sec. \ref{sec2}. Both quantities represent correlation 
energies gained by restoring (in an approximate way) the rotational 
and translational symmetries spontaneously broken by the mean field 
approximation. 
Moreover, to facilitate the analysis of the barriers 
heights, we have normalized energies to zero in the ground state.

In both nuclei the ground state is well deformed. Its quadrupole 
moment is $Q_2=8.3$ b ($\beta_2=0.18$) for $^{224}$Ra and $Q_2=14.1$ 
b ($\beta_2=0.27$)  for $^{238}$Pu. Small octupole deformation 
$Q_3=4.2$ b$^{3/2}$ ($\beta_3=0.14$) can be also found in the 
ground state of $^{224}$Ra. Fission valleys are characterized by 
a local decrease of the slope in the PES from the ground state 
towards scission. Fission paths can be found in the bottom of these 
valleys as to determine locally the lowest energy barriers. The direction 
corresponding to the slowest increase of the potential energy with 
deformation can be easily found along the reflection symmetric axis. 
This barrier is also plotted in Fig. \ref{barq2} with a green 
short-dashed line. At $Q_2=20-25$ b the barrier reaches a saddle 
point and then it slowly descends.
At larger elongation, from $Q_2=50$ b, the potential energy 
increases again producing a second hump of the barrier. At this 
stage, the fission valley turns into reflection asymmetric shapes 
and a second saddle point can be found around  $Q_2=55 - 60$ b with 
$Q_3=15$ b$^{3/2}$. This is  the typical scenario of fission in many 
heavy nuclei leading to asymmetric fission. Such valley is usually 
called ``elongated fission valley" \cite{war02, war05a} as the 
shapes of the nucleus along it are relatively stretched with a long 
neck coupling a typically spherical and a typically prolate-deformed 
nascent fragments. The value of the fission barrier height is 
around 10 MeV which is a value a little bit larger than values usually calculated
in the heavy actinides. In contrast to these nuclei \cite{war02, 
war05a, war05b} the barriers are extremely wide in light actinides.  
In $^{224}$Ra the potential energy oscillates around 10 MeV with 
increasing elongation and we have not been able to find a second 
turning point even for very large $Q_2$ values. The fission barrier 
of $^{238}$Pu finishes beyond $Q_2=100$ b. Very extended barrier 
cause long fission half-lives in all considered nuclei. Also the 
experimental branching ratio of spontaneous fission to $\alpha$ 
decay is very small in all of them  \cite{aud03a} and they are 
stable against fission.

In  Fig. \ref{map} one can also find a second valley on the PES  that
goes from  the ground state through the reflection asymmetric shapes with non-zero octupole 
moment. The huge octupole moment values obtained 
for small elongation suggests a large  asymmetry in the mass distribution.
As the saddle point is reached, the matter density distribution starts to 
resemble a molecular shape with a small sphere touching a larger one.  
The large spherical fragment has a number of protons and neutrons that is
consistent  with those of $^{208}$Pb. This observation points 
towards a clear relationship between  this valley and the phenomenon of 
CR. We will refer to this valley as ``hyper-asymmetric" or ``CR valley". Along 
the fission path in the bottom of this valley the elongation of the nucleus 
rises along with reflection asymmetry. Moreover, the fission 
path creates a straight line in the Figure~\ref{map}, as the growth of the 
quadrupole moment is proportional to the increase of the octupole moment.

\begin{figure*}
\includegraphics[width=1.0\columnwidth, angle=270]{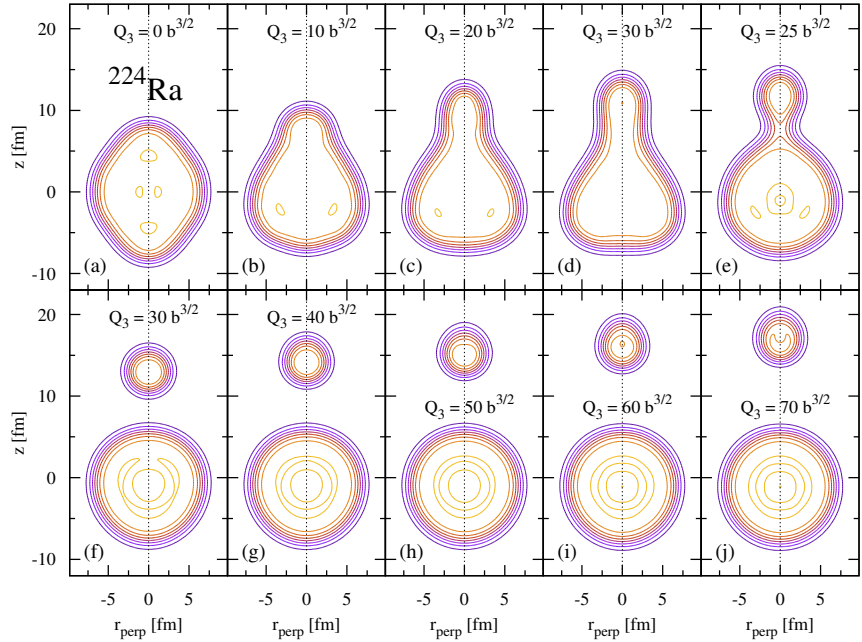}
\caption{\label{shapes224} (Color online) Shape evolution of $^{224}$Ra with increasing 
octupole moment $Q_3$. Panels (a)-(d) correspond to the up-going part of 
the fission path, panel (e)  to the short branch around saddle point and 
panels (f)-(j) correspond to the decreasing part of the fission path.}
\end{figure*}
\begin{figure*}
\includegraphics[width=1.0\columnwidth, angle=270]{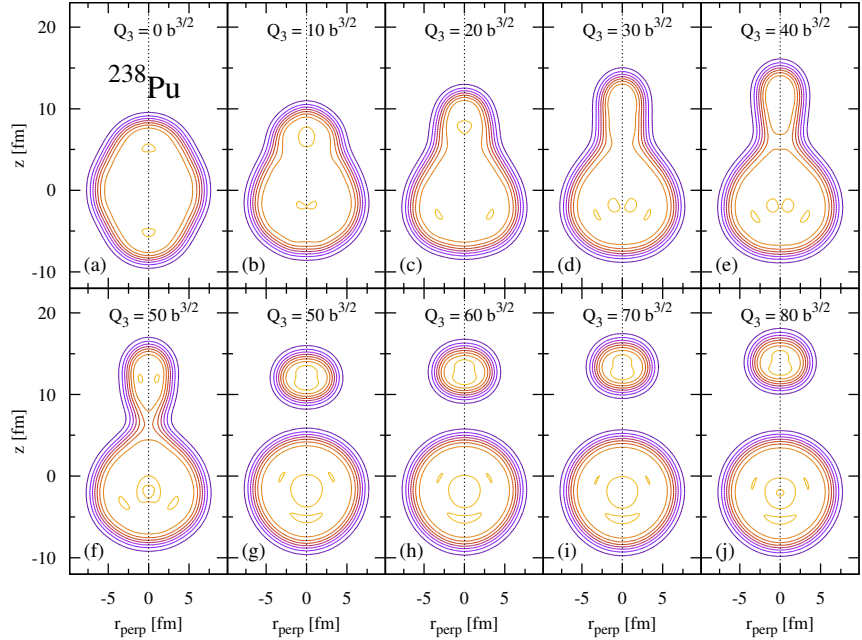}
\caption{\label{shapes238} (Color online)  The same as in Fig. \ref{shapes224} but 
for $^{238}$Pu. Panels (a)-(f) correspond to the up-going part of the 
fission path, and panels (g)-(j) correspond to the decreasing part of the fission path.}
\end{figure*}


In Fig. \ref{barq2} the hyper-asymmetric fission path is also plotted with a solid blue line as a 
function of the quadrupole moment. From this Figure, it is clear 
that the hyper-asymmetric barrier is much higher than the classical 
one. The potential energy grows very fast with deformation in the CR 
path up to around 25 MeV. Its height is extremely large in comparison 
with classical fission barrier. This implies very long half-lives for 
the decay along this channel (over $10^{10}$ s) and explains why
the CR path was ignored so far as the possible fission path.  The 
experimental evidences of CR, which is characterized by half-lives 
of the same order of magnitude, enforce to consider the 
hyper-asymmetric path as the possible exotic decay channel.

The evolution of the shapes of nuclei along the CR path from ground 
state to the saddle point is shown in Fig. \ref{shapes224} (a)-(e) for $^{224}$Ra 
and Fig. \ref{shapes238} (a)-(f) for $^{238}$Pu. One 
can see that a cluster of  nucleons is budding from the parent 
nucleus as elongation and asymmetry grow and already  at a modest octupole
deformation of $Q_3=20-30$ b$^{3/2}$ a neck starts to be clearly visible
in both cases.

Around $Q_2=20$ b a bifurcation can be found in the CR path of 
$^{224}$Ra. One of the branches goes towards large deformation 
parameters ($Q_2=45$ b, $Q_3=55$ b$^{3/2}$) with the energy reaching 
values over 40 MeV above the ground state. This path can not lead to 
fission, as the nucleus takes on it a cone-like shape (see Fig. \ref
{shapes224}d) without a well defined neck. The density profile 
corresponding to the shorter branch, presented in Fig. \ref
{shapes224}e, shows two nearly spherical fragments separated by a 
neck. This configuration is characterized by a hexadecapole moment 
which is substantially larger as seen in Fig. \ref{barq2}a. The same 
fission path bifurcation can be found also in $^{222}$Ra. In 
subsequent analysis we will consider only the second, shorter branch 
as the only relevant for the description of CR.

In the upper right corner of Fig. \ref{map} a distinct region of the 
PES is found, with the energy decreasing with increasing 
deformation.  In this part of the PES the system of nucleons is 
split into two fragments. It may be called ``fusion" valley in 
contrast to the first part which is commonly called ``fission" 
valley. The ``fusion" and ``fission" valleys are strongly 
correlated  with each other as they are linked in the same region of 
the deformation space and the mass splitting between fragments is 
similar in both cases. The minimum of the energy in this valley 
creates the descending branch of the CR fission barrier which is 
plotted with red dashed line in Fig. \ref{barq2}. In the upper 
panels of Fig. \ref{barq2} we observe the coincidence of the 
hexadecapole moments of both branches of the barrier at the saddle 
point.

The density distributions in the ``fusion" path are given 
in  Fig. \ref{shapes224} (f)-(j) for $^{224}$Ra and Fig. \ref
{shapes238} (g)-(j) for $^{238}$Pu. Some important information can 
be deduced from these plots. First, the system is built from two 
almost spherical fragments. The space between their surfaces is wide 
for at least a few femtometers and increase with $Q_2$ and $Q_3$. 
The heavier fragment is the doubly-magic spherical $^{208}$Pb after 
scission of  $^{224}$Ra or $^{210}$Pb in the case of  $^{238}$Pu. 
The lighter fragment may be slightly deformed (prolate or oblate).
Its shape is mostly determined by the shape of the ground state of
the corresponding nucleus as the Coulomb interaction with the
heavier fragment is not strong enough.
In the case of $^{30}$Mg emitted from $^{238}$Pu, the ground state 
is oblate ($\beta_2=-0.215$) \cite{mol95}. In the other nucleus, 
the spherical $^{14}$C isotope constitutes the lighter fragment of 
the CR from $^{224}$Ra. 

Once the system has split in two, the shapes of the fragments do 
not change significantly as they move apart and the increase of the 
total quadrupole and octupole momenta is a consequence of the 
increasing  distance between the fragments. Therefore the change in 
the potential energy after scission is mainly due to the decreasing 
of Coulomb repulsion and it should decline hyperbolically with the 
distance between the centers of fragments, which is roughly 
proportional to $Q_2$. Such behavior can be seen in Fig. \ref{barq2} 
close to the saddle point. However, for larger deformations we 
observe a departure from the expected behavior that calls for larger 
basis. Unfortunately, the use of larger basis can be problematic as 
a consequence of numerical instabilities in the evaluation of matrix 
elements due to finite computer accuracy. Those instabilities lead 
in some cases to strange behaviors in the energy preventing the use 
of very large basis (see also Appendix  \ref {appa} where the 
convergence of the energy is discussed). To avoid these 
difficulties, which are critical for the determination of half-lives 
in the WKB scheme, we have adopted an approximate strategy to be 
discussed in Sec. \ref{sec3b} below. The insufficient size of the 
basis also manifests in the matter distributions of the lighter 
fragment seen in  panels (i) and (j) of Fig. \ref{shapes224} where 
an unnatural stretching towards large $z$ values can be noticed.

The solution of the HFB equation often depends on the  nuclear 
matter distributions of the initial wave function used in the 
iterative  procedure. In many regions of the PES, especially close 
to the scission line, two solutions may be obtained for the same 
constrains. If the calculation begins with a compact shape of the 
nucleus, the final solution will have similar properties. If a 
configuration with two separated fragments is chosen as the  initial 
condition, again a solution with similar properties will be found. 
When the same constrains are put on the system both results will 
have the same quadrupole and octupole momenta but they may have 
different higher multipolarities as well as energy. Since that, 
``fusion" valley extends towards ground state much further than it 
is shown in Fig. \ref {map}. It covers the area around ``fusion" 
path in the $Q_2-Q_3$ deformation space. Its part is hidden below 
the ``fission" valley shown in Fig. \ref{map}. In this Figure we 
have marked both the fission paths and  ``fission" valley but not 
the whole ``fusion" valley.

\subsection{Tracking fission paths as a function of octupole moment 
\label{sec3b}}

\begin{figure}
\includegraphics[width=0.7\columnwidth, angle=270]{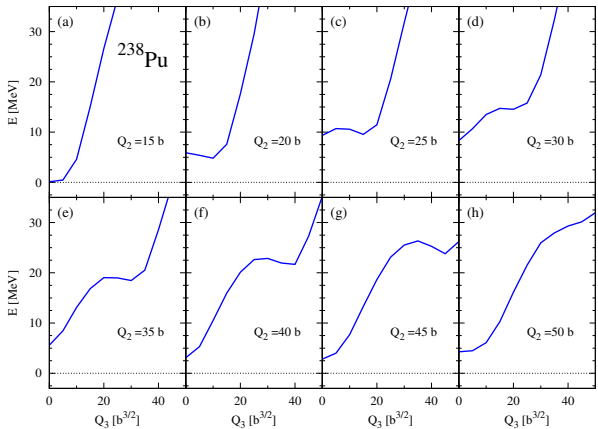}
\caption{\label{cutq2} (Color online) Potential energies of $^{238}$Pu as a function 
of the octupole moment $Q_3$ for several values (shown in each panel) 
of the  quadrupole moment $Q_2$. }
\end{figure}

\begin{figure}
\includegraphics[width=0.7\columnwidth, angle=270]{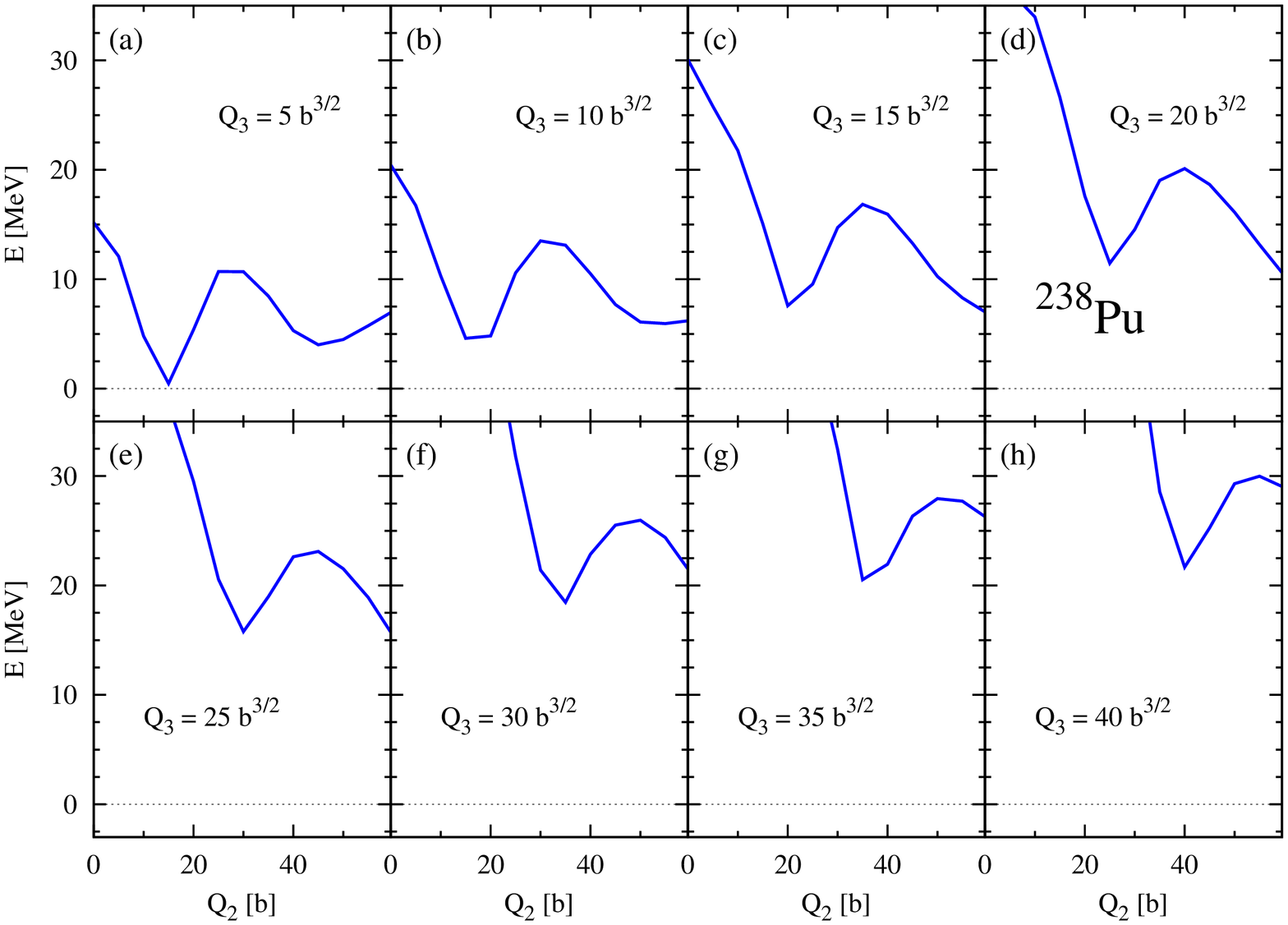}
\caption{\label{cutq3} (Color online) Potential energies of $^{238}$Pu as a function 
of the quadrupole moment $Q_2$ for several values (shown in each panel)
of the octupole moment $Q_3$. }
\end{figure}

Tracking the hyper-asymmetric fission path in the PES is a difficult task from a 
numerical standpoint. Usually, the  fission path is determined by 
searching for the local minima of the energy along cuts of constant 
$Q_2$. This method could also be applied to the CR path as it is 
shown in Fig. \ref {cutq2}, where the potential energies of the 
$^{238}$Pu nucleus are plotted as a function of $Q_3$ for fixed 
values of $Q_2$. It is clear that local minimum corresponding to the 
hyper-asymmetric fission can be determined in most of the cases, 
usually at higher energies than the minimum of the classical fission 
observed at $Q_3=0$ b$^{3/2}$. However, in many nuclei there are 
certain $Q_2$ values where a plateau is observed instead of a well 
defined minimum (e.g. for $Q_2=30$ b in $^{238}$Pu in  Fig. \ref 
{cutq2}d). This problem can be solved by using an alternative choice 
of coordinate to describe the formation of the daughter nuclei. As 
it has been mentioned before, in the CR path $Q_2$ is roughly 
proportional to $Q_3$ and therefore the octupole moment can be also 
used as the leading coordinate. The potential energy for fixed $Q_3$ 
as a function of $Q_2$ is plotted in Fig. \ref{cutq3}. Here the hyper-asymmetric 
valley is clearly visible at every point and it is trivial to 
determine local minima there and track the fission path. The 
octupole moment can also be used as the driving coordinate to 
determine half-lives in the WKB approximation as already described 
in the Sec. \ref{sec2}. We conclude that the octupole moment is 
better suited than the quadrupole moment to describe the CR paths in 
the PES and therefore we will use it as a leading coordinate in the 
following discussion. 

\begin{figure*}
\includegraphics[width=1.0\columnwidth,angle=270]{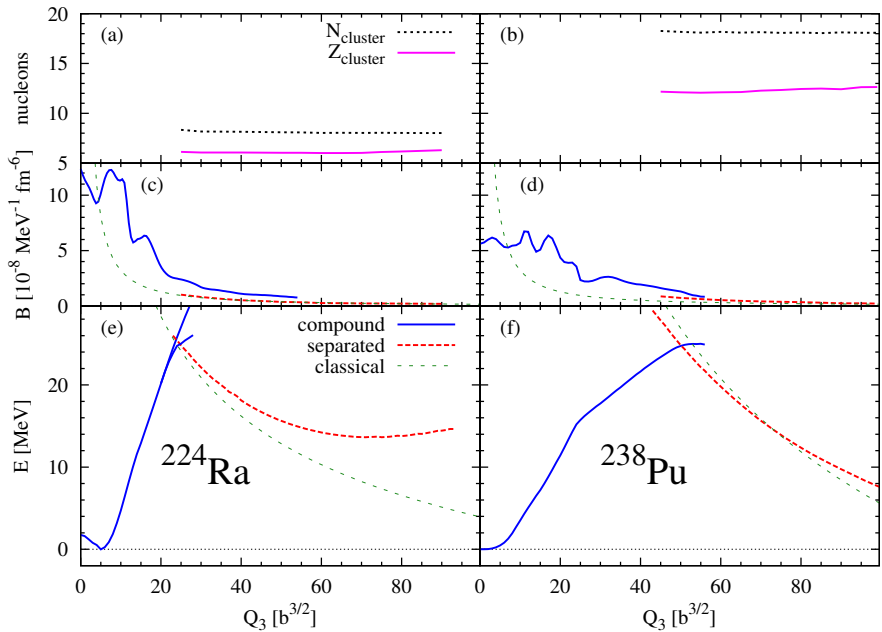}
\caption{\label{barq3} (Color online) Hyper-asymmetric fission barriers in 
$^{224}$Ra (left) and $^{238}$Pu (right) as a function of the octupole moment 
$Q_3$ (lower panels). Approximate Coulomb repulsion energies [Eq. (\ref{coulomb})] for corresponding clusters
are also plotted. In the middle panels, the mass parameter 
$B(Q_3)$ calculated in a microscopic way is plotted. 
In addition,  the classical value [Eq. (\ref{bq3})] corresponding
to two separate fragments is also given. In the upper panel, the 
number of nucleons in clusters is given as a function of the octupole
moment $Q_3$.}
\end{figure*}

The profiles of the CR path in $^{224}$Ra and $^{238}$Pu, presented 
already as a a function of quadrupole moment in Fig. \ref{barq2},  
are plotted now as a function of octupole moment in Fig. \ref{barq3} 
(e) and (f). Initially, the energy increases with increasing octupole moment in 
an almost quadratic fashion from the ground-state which may be 
refection symmetric ($^{238}$Pu) or asymmetric ($^{224}$Ra). The 
slope of energy decreases  when 
approaching to the top of the barrier. At some point the branch with 
two fragments becomes the lowest energy solution with an energy 
which is essentially the Coulomb repulsion of the fragments 
expressed as a function of the octupole moment of the two fragments. 
The Coulomb repulsion energy can be very well be approximated by the 
classical value corresponding to two uniformly charged spheres:

\begin{equation}
V(Q_3)=V_{Coul}-Q=e^{2}\frac{Z_{1}Z_{2}}{R}-Q.
\label{coulomb}
\end{equation} 
Asymptotically, the total energy tends to the $Q$ value of the 
reaction that can be  extracted from the experimental binding 
energies \cite{aud03b}. In the above expression $R$ represents the 
distance between the centers of mass of the fragments. The 
connection between the variable $R$ and the octupole moment $Q_{3}$ 
is obtained in a simple geometrical way when the two fragments are 
spherical or when two 
point masses are considered
\begin{equation}
Q_{3}=f_{3}R^{3},
\label{qr}
\end{equation}
where
\begin{equation}
f_{3}=\frac{A_{1}A_{2}}{A}\frac{(A_{1}-A_{2})}{A}
\label{f3}
\end{equation}
is given in terms of the total mass number $A$, and the mass numbers 
of each of the fragments $A_{1}$, and $A_{2}$. In Fig. \ref{barq3} 
(e) and (f) we observe that around the saddle point both the HFB and 
the approximate Coulomb repulsion energy of Eq. (\ref{coulomb}) 
coincide with a noticeable agreement of the order of 2 MeV. Small 
differences can be mainly attributed to the excitation of the 
lighter fragment in the presence of the Coulomb field of the heavy 
mass residue as well as to the deformation of the emitted cluster 
that can be different from the one of its ground-state. As it was 
already mentioned in Sec. \ref{sec3a}, at larger $Q_3$ values the 
HFB energy results are more affected by the finite size of the basis 
used and therefore they lie at an energy higher than the one of an 
infinite basis calculation. As this range of octupole moments is 
very relevant for the determination of half-lives in the WKB 
framework, we  will use the approximate expression of Eq. (\ref
{coulomb}) in the calculation of half-lives instead of the HFB energy.

The collective mass $B(Q_{3})$ linked to the octupole moment is also 
plotted in Fig. \ref{barq3} (c) and (d).  The collective mass of the 
compound system computed microscopically substantially differs from 
the semi-classical value given by the reduced mass of the two 
fragments $\mu=m_{n}\frac{A_{1}A_{2}}{A_{1}+A_{2}}$ (a quantity 
connected to the kinetic energy for the coordinate $R$) but written 
in terms of $Q_{3}$ 
\begin{equation}
B(Q_{3})=\frac{\mu}{9Q_{3}^{4/3}f_{3}^{2/3}}.
\label{bq3}
\end{equation}
This quantity derived from ATDHFB model in Eq. (\ref{batdhfb}) 
varies considerably when the nucleus is stretched out. That is a 
consequence of the strong dependence of collective mass on the 
single particle effects that show up during the development of the 
neck. After scission the microscopic collective mass $B(q_{3})$ is 
very close to the classical value, as expected.

In  panels (a) and (b) of Fig. \ref{barq3} we have shown the numbers 
of protons and neutrons of the lighter fragment after scission in 
the CR path. In this way the cluster emitted in the hyper-asymmetric 
fission can be identified. In $^{224}$Ra it corresponds exactly to 
the experimentally observed cluster $^{14}$C. The PES of $^{238}$Pu 
indicates $^{30}$Mg as a potential cluster. This is one of the 
clusters observed in the decay of this nuclide ($^{28,30}$Mg and 
$^{32}$Si). 

We would like to point out and important aspect of tracking the 
fission path after the scission point. In the laboratory it is not 
possible to transfer nucleons  between the daughter nuclei once the 
fragments are created. We have checked that the numbers of neutrons 
and protons are usually constant in the minimum of the ``fusion" 
valley, although they may differ slightly from integer numbers. 
Imposing given fragment masses will lead to configurations with 
higher energies. The ``fusion" paths for those systems  with mass 
asymmetry differing  by a few nucleons from the one corresponding to 
the minimum energy configuration run parallel to the minimum energy 
path and they reach the scission point at almost the same position 
in the $Q_3-Q_4$ deformation space, i.e. in the saddle, with similar 
energy. A tiny instability around the saddle may lead to alternative 
choice of cluster configuration. Length of the fission barrier 
corresponding to each possible nascent fragment would determine 
which one will be observed in experiment. Further detailed 
investigations should be performed using additional constraint on 
the number of nucleons of each fragment.

\subsection{Scission point transition from compound nucleus to two separated fragments.}
\label{sec3c}

\begin{figure*}
\includegraphics[width=1.5\columnwidth, angle=270]{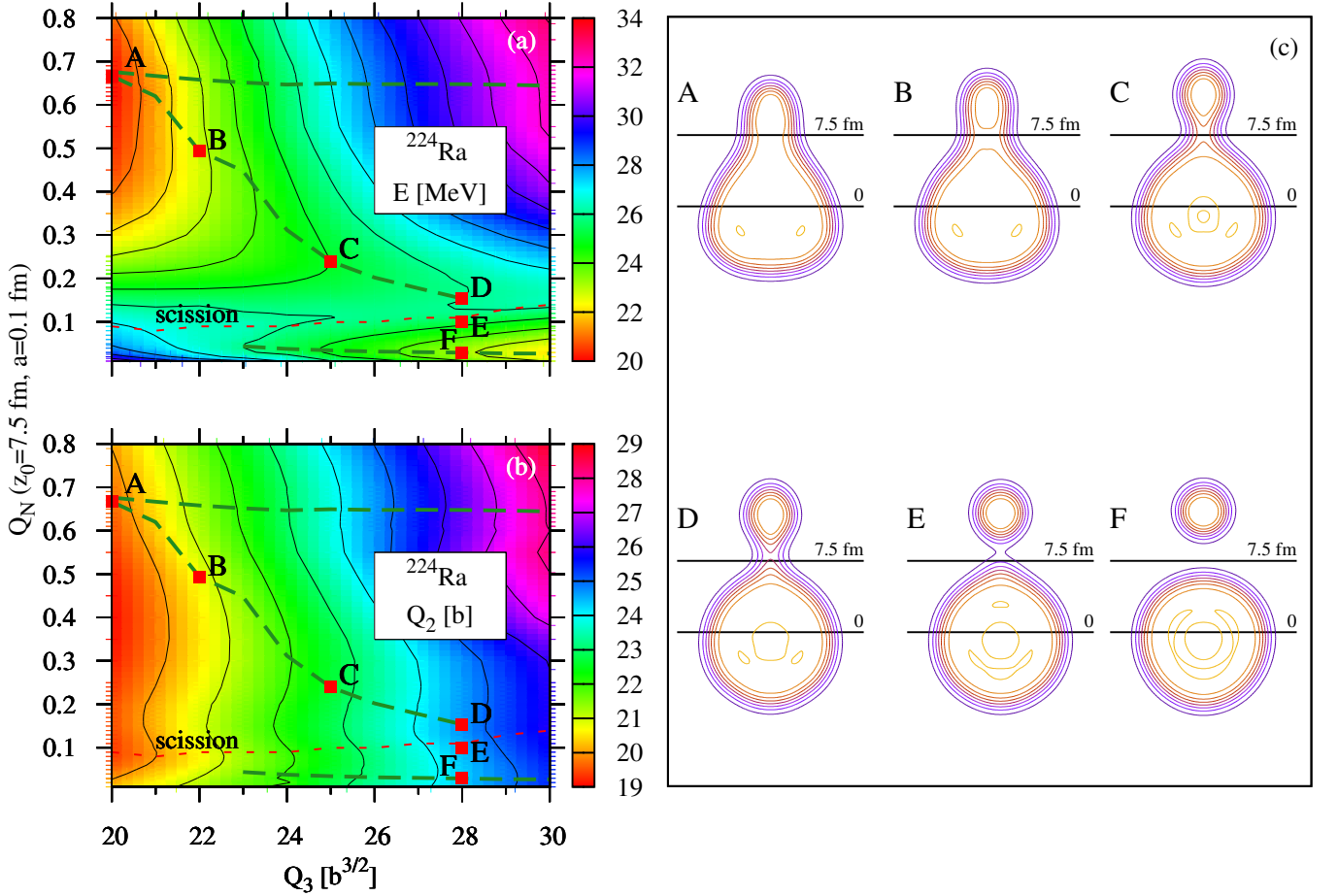}
\caption{\label{scission224} (Color online) (a) Saddle region of the PES and 
(b) quadrupole moment $Q_2$ plotted as a function of octupole moment $Q_3$ 
and neck parameter $Q_N$ in $^{224}$Ra. Green dashed lines are fission paths and  
red short-dashed line is scission line. 
(c) Contour plots of density distributions for some deformations marked in 
panel (a). Equidensity lines are plotted every 0.02 fm$^{-3}$.}
\end{figure*}

\begin{figure*}
\includegraphics[width=1.5\columnwidth, angle=270]{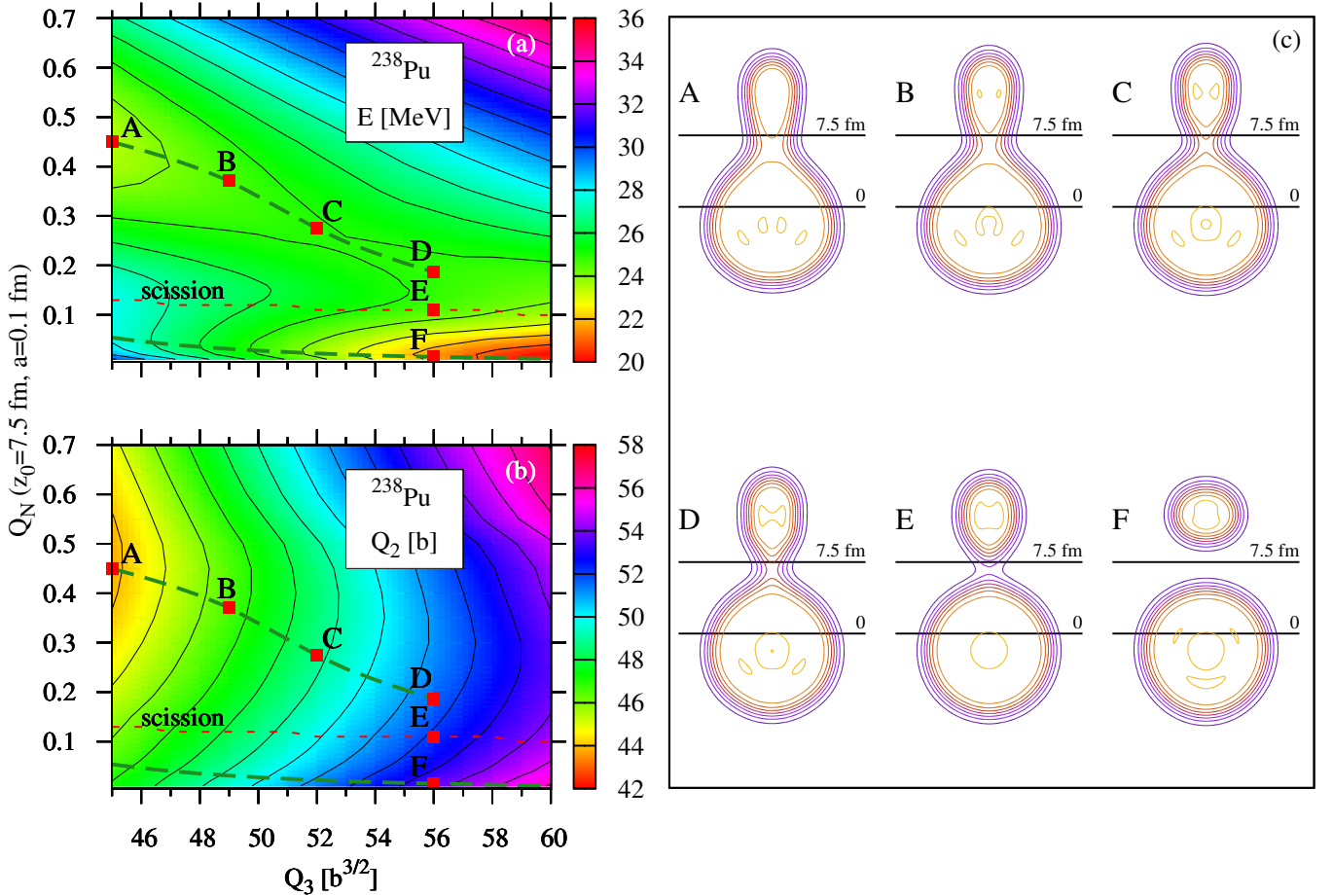}
\caption{\label{scission238} (Color online)  The same as in Fig. \ref{scission224} but for the
 $^{238}$Pu nucleus.}
\end{figure*}

\begin{figure}
\includegraphics[width=0.7\columnwidth, angle=270]{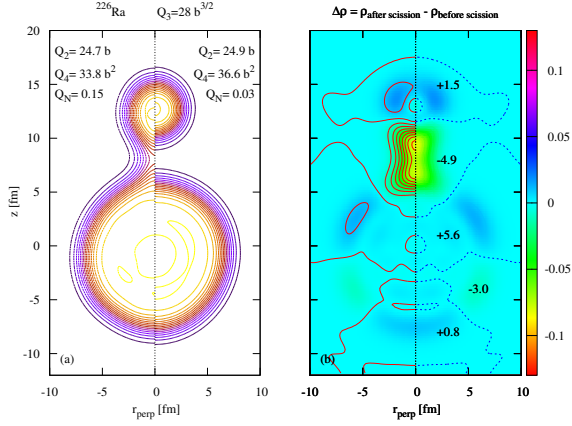}
\caption{\label{densdif224} (Color online) In the panel (a) a comparison of the density distribution 
of  $^{224}$Ra before and after scission at $Q_2=28$ b$^{3/2}$ is shown.  
Equidensity lines are plotted every 0.01 fm$^{-3}$. 
In panel (b) the
differences between the two matter density distributions of the
 panel (a) are plotted. In this plot, contour lines 
in the left part are plotted every 0.01 fm$^{-3}$, whereas on
the right hand side only for $\Delta\rho=0$. The number of nucleons shifted 
to (+)   or from (-) the marked region are also given. }
\end{figure}

\begin{figure}
\includegraphics[width=0.7\columnwidth, angle=270]{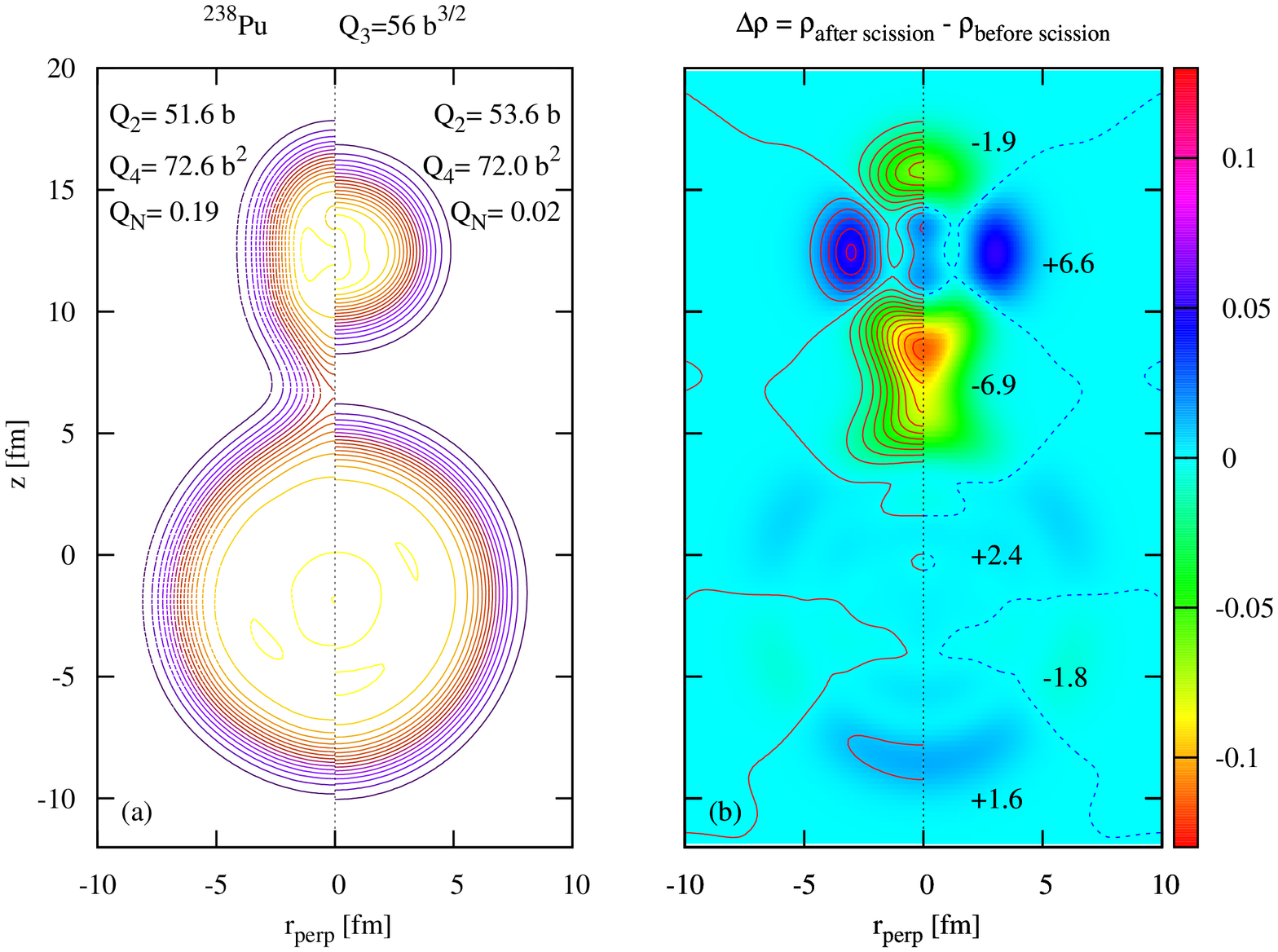}
\caption{\label{densdif238} (Color online) The same as in Fig. \ref{densdif224} but 
for the $^{238}$Pu nucleus at $Q_3=56$ b$^{3/2}$. }
\end{figure}

Two independent branches are clearly visible in the CR fission 
barriers of Figs. \ref{barq2} and \ref{barq3}. As described in Sec. 
\ref{sec3b} they differ substantially in the shapes of the nucleus 
corresponding to each of them. In the first, up-going part of the 
barrier, called ``fission" path, the shape corresponds to the one of 
a  compound nucleus [Figs. \ref{shapes224}(a)-(e) and \ref 
{shapes238}(a)-(f)]. For the deformations around the ground state 
the corresponding shape is not too distant from the ellipsoid and 
therefore we can say that the nucleus takes a compact shape. 
However,  for large deformations a neck can  be clearly 
distinguished on this branch and the density distribution of the 
nucleus is of molecular type. A completely different type of shape 
is obtained on the down-slope side of the barrier, called ``fusion" 
path [Figs. \ref{shapes224}(f)-(j) and \ref {shapes238}(g)-(j)]. Two 
well separated nuclei can be observed there as the matter density in 
the region between them goes to zero and the  shortest distance 
between  the nuclear surfaces of the two fragments  is at least of a 
few femtometers.

In spite of the different shapes of the  density profiles along the 
two branches, many nuclear properties are similar in both of them at 
the top of the barrier. For instance, comparable values of the 
quadrupole, octupole and hexadecapole moments can be found there. 
The density distribution before scission is close to the one after 
separation and only important differences can be found in the neck 
region. Moreover, the energies also  have similar values and we can 
easily find in Fig. \ref{barq3} a crossing point where the potential 
energy on both branches is the same for some value of $Q_3$. A first 
and rough approximation could be to consider this as the scission 
point. This assumption can be used to get a quite reasonable 
estimation of the size of the barrier  and fission half-lives. 
Nevertheless, a more precise analysis shows that the shape of the 
system is clearly different in both branches and none of them can be 
considered as two touching fragments. 

The passage from a compact shape to a two fragments one can not be 
treated as an instant transition at the crossing point. Some energy 
barrier, not seen clearly in the PES spanned in the $Q_2-Q_3$ space, 
exists between the  ``fission" and the ``fusion" path. These two 
constraints are not sufficient to describe the continuous path 
connecting both branches. In such a path, the nuclear density in the 
neck would decrease gradually to zero and then two fragments would 
be disengaged. The relevant parameter along this path is the neck 
parameter \cite{ber90,war02} defined through the mean value of the 
operator 

\begin{equation}
\hat Q_N=\exp\left(\frac{(z-z_0)^2}{a^2}\right) \,\,.
\end{equation}

The value of the neck parameter roughly corresponds to the number of 
nucleons in a slice perpendicular to the $z$ axis, centered at the  
position $z_0$ and of width $a$. In the present case we have chosen 
$a=0.1$~fm, which gives us a sufficiently thin slice, and $z_0=7.5$
~fm which corresponds to the position of the neck. The neck 
parameter is correlated with the hexadecapole moment, a quantity 
that has been used routinely in fission calculations \cite{ber90} to 
study the scission process,  but the neck parameter  is more suited 
to drive the system through scission when $z_0$ and $a$ are chosen 
conveniently. The quantity $Q_N$ never goes to zero in any physical 
situation because of the non-vanishing tail of the nuclear density 
distribution but it can be arbitrarily small if the  slice is 
properly located in the region between the two separated fragments.

In panel (a) of Figs. \ref{scission224} and \ref{scission238} the 
PES of $^{224}$Ra and $^{238}$Pu are plotted, 
respectively, as a function of the octupole moment $Q_3$ and the 
neck parameter $Q_N$. In these plots, we only show the relevant 
region around the top of the barriers. The minima of the valleys on 
this surface are marked by green dashed lines. The ``fission" path 
goes from $Q_3=20$ b$^{3/2}$, $Q_N=0.65$ to $Q_3=28$ b$^{3/2}$, 
$Q_N=0.18$  in $^{224}$Ra and from $Q_3=45$ b$^{3/2}$, $Q_N=0.45$ to 
$Q_3=55$ b$^{3/2}$, $Q_N=0.20$  in $^{238}$Pu. The ``fusion"  path 
is marked by an almost horizontal line with neck parameter in the 
range from  $Q_N=0.02$ to $Q_N=0.05$ in both nuclei. In panel (a) of 
Fig. \ref{scission224} a horizontal line at $Q_N=0.65$ is also 
shown. It corresponds to the branch of the fission path in $^{224}$Ra 
that goes up in energy and that shows shapes that do not develop 
a sizable neck. The red short-dashed line around $Q_N=0.10$ is the 
scission line describing these configurations where the density in 
the neck region goes below 0.4 fm$^{-3}$. It lies along the ridge on 
the energy surface separating  ``fission" and ``fusion" valleys. 
Thanks to the use of  the neck parameter both valleys are linked 
in a continuous way along the whole scission line and there is no 
sudden energy change.

In panel (b) of  Figs. \ref{scission224} and \ref{scission238} the 
quadrupole moment of the nucleus is plotted in the same space of 
deformations as in panel (a) of the same figures. We observe how the 
quadrupole moment increases monotonically  with increasing octupole 
moment. The variations with the neck parameter are much smaller, 
though. This explains why the quadrupole moment does not provide a 
quantity sensitive  enough for the detailed description of the 
rupture of nucleus in two pieces. The hexadecapole moment is also not 
sensitive for changes of the neck parameter and it varies by not more than 4 b$^2$
for the fixed octupole moment in the configurations considered in the Figures.
Larger monotonic increase of $Q_4$ with $Q_3$ can be observed.

Finally, in panel (c) of Figs. \ref{scission224} and \ref{scission238} 
the matter density distribution at the different stages
of the scission process [marked by the letters A, B, C, D, E, and F in
panels (a) and (b)] is shown. Following the points at the ``fission" path marked as 
A, B, C and D a reduction of the neck parameter can be noticed. The 
neck  becomes thinner and  a decrease of the  nuclear density up to  
half of the bulk value in the configuration D can be observed. Between the
configurations D, E and F the scission process takes place and the shape
of the  lighter fragment evolves from prolate in D to spherical or oblate in F. It
is also interesting to notice that shapes D, E and F have essentially
the same octupole moment and a very similar quadrupole moment.

We also observe in panel (a) of Figs. \ref{scission224} and \ref
{scission238} that  the crossing point of the two branches 
(``fission" and ``fusion") at $Q_3=50$ b$^{3/2}$ for $^{238}$Pu 
and at $Q_3=25$ b$^{3/2}$ for $^{224}$Ra are well separated as they 
correspond to  different values of $Q_N$. It is now clear that to 
pass directly from one configuration to the other it is necessary to 
climb the ``neck barrier" which is  over 1 MeV high, although the 
energy of both the ``fission" and ``fusion" paths is the same.
From these plots it becomes clear that it is energetically 
preferable to follow the ``fission" path to the very end, where the 
neck is very thin (see the shape of the nucleus at the point D) and 
there is no barrier separating the nucleus from the scission line, 
than to climb the ``neck barrier". The subsequent evolution of the 
shape of the nucleus should follow the direction corresponding to 
the maximal decrease in energy (the gradient direction). In fact, it 
means that the neck parameter should decrease rapidly almost without 
change of the octupole moment until it reaches the bottom of the 
``fusion" valley. In this way the nucleus  takes first the shape 
corresponding to the configuration E and then the one corresponding to 
the configuration F at the  ``fusion" path. At this point, the direction 
of  largest energy slope corresponds to the   ``fusion" path. The 
nuclear profile E really represent a scission configuration with two 
nuclei that have common only tail of the density distribution below 
the value of 0.04 fm$^{-3}$.
 
In this way we can follow step by step the sequence of changes of 
the nuclear shape around the scission point without loosing 
continuity. From the present analysis we can obtain a precise 
outline of the CR fission barrier in the actinides.  The first fragment
of the barrier is 
built from the whole ``fission" branch where energy increase with 
deformation. After its termination a 
rapid decrease of the energy takes place without changes in 
quadrupole or octupole moment until the ``fusion" path is reached. 
Finally further decrease of energy with increasing deformation is 
observed in the ``fusion" path.  Although such shape seems to 
lose continuity in the plot of the energy barrier as a function of 
quadrupole or octupole moment (see Figs. \ref{barq2} and \ref{barq3}),  
it is continuous in the space spanned on the neck 
parameter.

Let us now take a closer look to the shapes of nucleus at the end of 
the ``fission" path and just after reaching the ``fusion" path. i.e. 
from points D and F in Figs. \ref{scission224} and \ref{scission238}.
These configurations differ in energy by a few MeV 
and they are distinguished by their matter distribution at  the 
neck. Nevertheless, the distribution of nuclear matter is quite 
similar in both cases. In Fig. \ref{densdif224}a the density 
distributions of the nucleus  $^{224}$Ra before (on the left hand 
side) and after scission (on the right hand side) are shown. Both 
nuclear system have got the same octupole moment $Q_3=28$ b $^{3/2}$ 
and similar quadrupole and hexadecapole moments ($Q_2=24.7$ b, 
$Q_4=33.8$ b$^2$ for compound shape and $Q_2=24.9$ b, $Q_4=36.6$ b$^2$ 
after scission). First of all we notice that in both cases the 
larger fragment is the same spherical doubly-magic $^{208}$Pb 
nucleus. It is well developed before the rupture of the neck and 
only small transfer of nuclear matter can be seen within this part 
of the system. The smaller fragment $^{14}$C is also present before 
scission. Its central part is well separated from the heavier 
fragment and its spherical shape is almost developed before rupture 
of the neck, although small prolate deformation can be noticed. 
Nuclear density in the neck is lower than the bulk nuclear density 
and goes down to 0.09 fm$^{-3}$ in the molecular configuration. The 
distance between the centers of the two incipient fragments before 
scission is the same as between the separated fragments after 
scission.

Further analysis of the matter distribution at the scission point 
configuration requires of Fig. \ref{densdif224}b where the 
differences between the density distributions depicted in panel (a) 
of the same Figure are shown. Only approximately 4.9 nucleons are 
transferred from the neck to the 
fragments. Small shifts of nuclear matter can be also observed 
within each of the fragments. The 
heavy fragment is not changing in a remarkable way. 
The light fragment is relatively more 
affected by the displacement of nuclear matter and changes of its 
shape.

Similar conclusion  can be deduced for $^{238}$Pu from Fig. \ref{densdif238}. 
On both sides of panel (a) the nucleus has the same 
octupole moment $Q_3=56$ b$^{3/2}$ and similar quadrupole and 
hexadecapole moments ($Q_2=51.6$ b, $Q_4=72.6$ b$^2$ for molecular 
shape and $Q_2=53.6$ b, $Q_4=72.0$ b$^2$ for two fragments). Changes 
in the density distribution in the cluster region are slightly 
larger than in $^{224}$Ra, since a change of deformation in the 
lighter fragment from prolate to oblate can be noticed. The $^{30}$Mg nucleus
corresponding to the lighter fragment has a ground state which is 
very soft against changes of its quadrupole deformation \cite{cea}. 
Before scission the lighter fragment of the compound nucleus is 
stretched to have a prolate shape whereas after scission it takes 
its oblate ground state deformation ($\beta_2=-0.222$ \cite{mol95}). 

The constraint on the neck parameter allows for a detailed analysis 
of the scission point configuration. In consequence a continuous 
fission path going from the compound nucleus to the two separated 
fragments could be determined. The precise localization of the saddle point 
can be also settled using this parameter.

\section{Result for actinide nuclei \label{sec4}}

\begin{figure*}
\includegraphics[width=1.9\columnwidth]{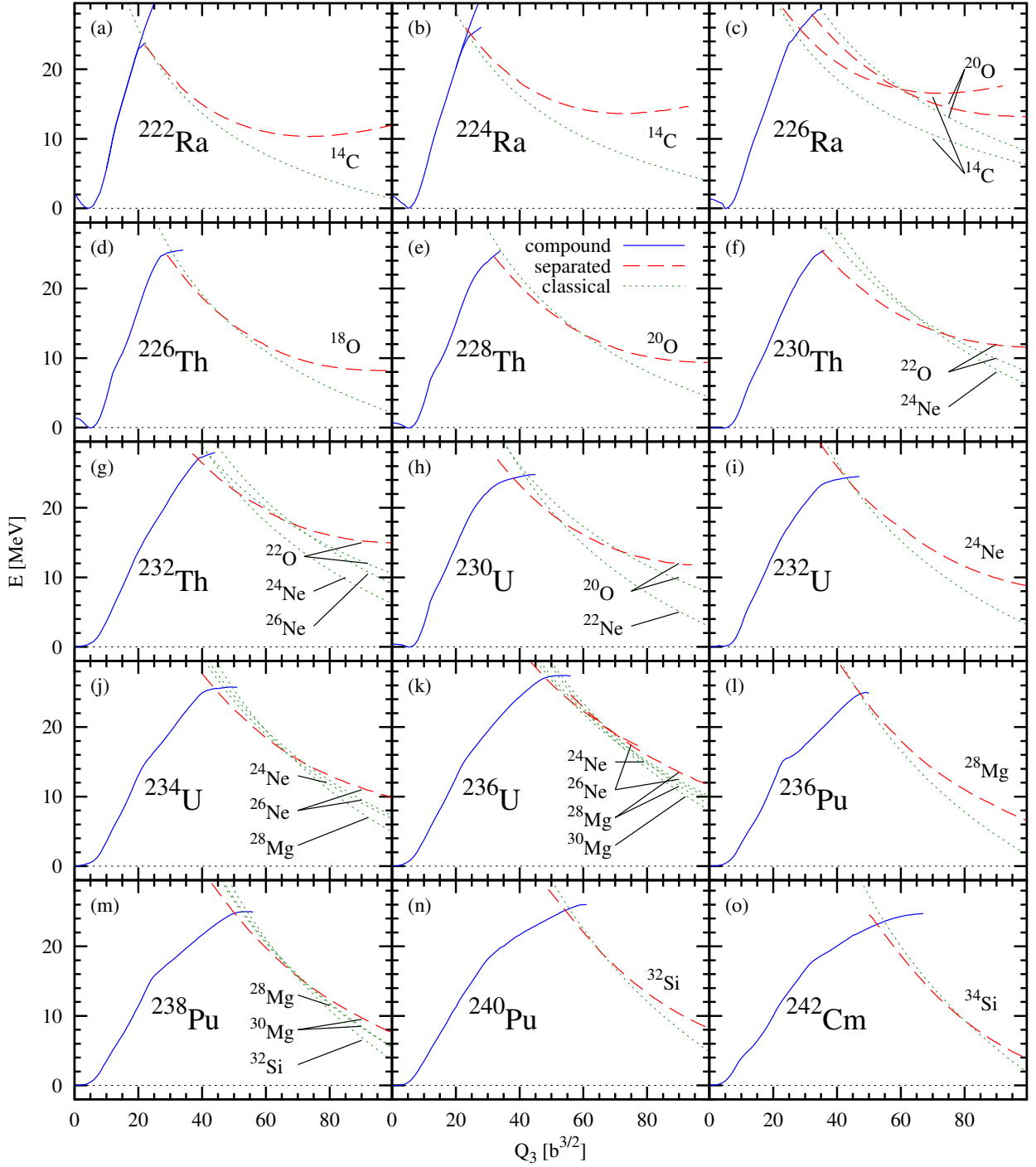}
\caption{\label{allbar} (Color online) Hyper-asymmetric fission barriers 
for all considered isotopes as a function of octupole moment $Q_3$. 
Fragment of the barrier with compound nucleus is marked with a blue solid 
line and with two separated fragments with red dashed line. Green dotted 
lines show classical Coulomb energy for two fragments. The corresponding 
clusters are indicated for solutions after scission.}
\end{figure*}

\begin{figure*}
\includegraphics[width=1.1\columnwidth,angle=270]{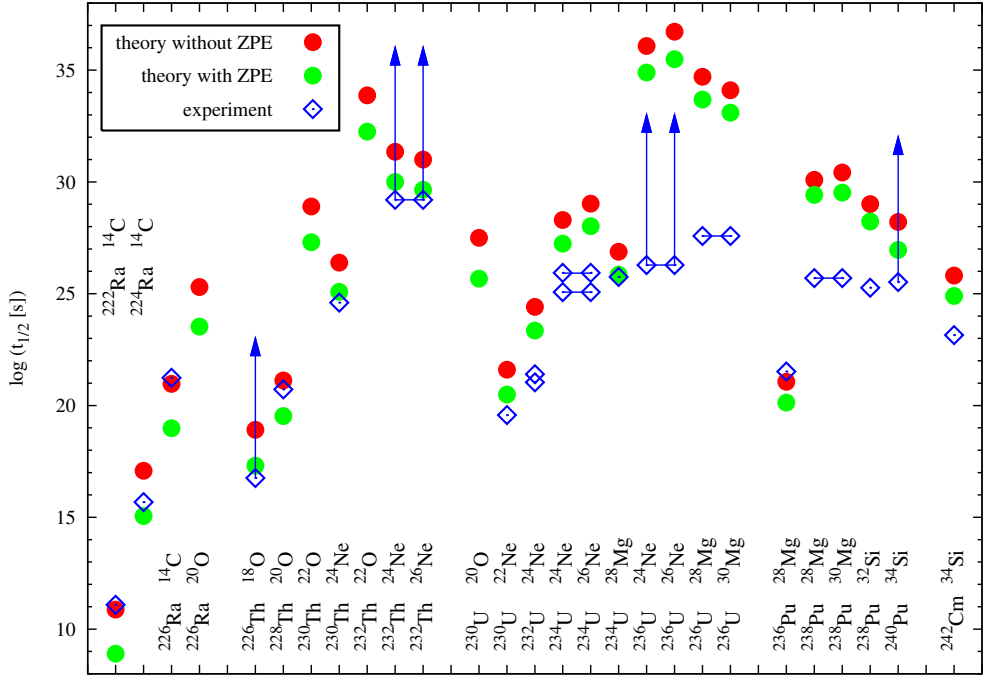}
\caption{\label{halflives} (Color online) Half-lives for cluster emission of various 
isotopes and various clusters. Blue diamonds show the experimental 
half-lives. Arrows indicate low experimental limit. Connected diamonds are for 
experimental values for two clusters. If experimental data from different 
experiments exceed differ by more than 0.3, the extreme values are indicated. 
 }
\end{figure*}

\begin{table}
\renewcommand{\arraystretch}{1.2}
\begin{tabular}{cccccc}
\hline
\hline

emitter     & cluster   & residue    &$\log(t_{1/2}^\mathrm{HFB}$[s]) &$\log(t_{1/2}^\mathrm{exp}$[s]) & ref.\\
\hline
\hline
 $^{222}$Ra & $^{14}$C  & $^{208}$Pb &  8.90                  &    11.01 & \cite{pri85} \\
            &           &            &                        &    11.09 & \cite{hou85} \\
            &           &            &                        &    11.22 & \cite{hus91} \\
\hline
 $^{224}$Ra & $^{14}$C  & $^{210}$Pb & 15.06                  &    15.86 & \cite{pri85} \\ 
            &           &            &                        &    15.68 & \cite{hou91} \\            
\hline
 $^{226}$Ra & $^{14}$C  & $^{212}$Pb & 18.98                  &    21.19 & \cite{hou85} \\
            &           &            &                        &    21.24 & \cite{bar86} \\ 
            &           &            &                        &    21.34 & \cite{wes90} \\     
\hline
 $^{226}$Ra & $^{20}$O  & $^{206}$Hg & 23.53                  &      -   &              \\
\hline
\hline
 $^{226}$Th & $^{18}$O  & $^{208}$Pb & 17.31                  & $>$16.76 & \cite{bon01} \\
\hline

 $^{228}$Th & $^{20}$O  & $^{208}$Pb & 19.53                  &    20.72 & \cite{bon93} \\
\hline

 $^{230}$Th & $^{22}$O  & $^{208}$Pb & 27.30                  &      -   &              \\
\hline
 $^{230}$Th & $^{24}$Ne & $^{206}$Hg & 25.08                  &    24.60 & \cite{bon99} \\
\hline

 $^{232}$Th & $^{22}$O  & $^{210}$Pb & 32.25                  &      -   &              \\ 
\hline
 $^{232}$Th & $^{24}$Ne & $^{208}$Hg & 30.00                  &  $>$29.20 & \cite{bon95}\\

 $^{232}$Th & $^{26}$Ne & $^{206}$Hg & 29.65                  &          &               \\
\hline
\hline
 $^{230}$U  & $^{20}$O  & $^{210}$Po & 25.67                  &    -   &              \\
\hline
 $^{230}$U  & $^{22}$Ne & $^{208}$Pb & 20.49                  &    19.57 & \cite{bon01} \\

\hline
 $^{232}$U  & $^{24}$Ne & $^{208}$Pb & 23.35                  &    21.04 & \cite{bar85} \\
            &           &            &                        &    20.40 & \cite{bon90} \\
            &           &            &                        &    20.39 & \cite{bon91} \\

\hline

 $^{234}$U  & $^{24}$Ne & $^{210}$Pb & 27.24                  &    25.07 & \cite{wan87} \\
 $^{234}$U  & $^{26}$Ne & $^{208}$Pb & 28.02                  &    25.25 & \cite{moo89} \\
            &           &            &                        &    25.30 & \cite{tre89}\\
            &           &            &                        &    25.93 & \cite{bon91} \\
            &           &            &                        &    25.89 & \cite{bon91} \\ 
\hline

 $^{234}$U  & $^{28}$Mg & $^{206}$Hg & 25.85                  &    25.54 & \cite{wan87} \\
            &           &            &                        &    25.75 & \cite{moo89} \\
            &           &            &                        &    25.54 & \cite{tre89} \\
            &           &            &                        &    25.75 & \cite{bon91} \\

\hline

 $^{236}$U  & $^{24}$Ne & $^{212}$Pb & 34.89                  & $>$26.28 & \cite{tre89} \\

 $^{236}$U  & $^{26}$Ne & $^{210}$Pb & 35.49                  &          &              \\
\hline
 $^{236}$U  & $^{28}$Mg & $^{208}$Hg & 33.68                  & $>$26.28 & \cite{tre89} \\
 $^{236}$U  & $^{30}$Mg & $^{206}$Hg & 33.10                  &    27.58 & \cite{tre94} \\
\hline
\hline

 $^{236}$Pu & $^{28}$Mg & $^{208}$Pb & 20.13                  &    21.67 & \cite{ogl90} \\
            &           &            &                        &    21.52 & \cite{hus95} \\
\hline
 $^{238}$Pu & $^{28}$Mg & $^{210}$Pb & 29.42                  &    25.70 & \cite{wan89} \\

 $^{238}$Pu & $^{30}$Mg & $^{208}$Pb & 29.52                  &          &              \\
\hline
 $^{238}$Pu & $^{32}$Si & $^{206}$Hg & 28.23                  &    25.27 & \cite{wan89} \\
\hline

 $^{240}$Pu & $^{34}$Si & $^{206}$Hg & 26.96                  & $>$25.52 & \cite{bon99}\\
\hline
\hline

 $^{242}$Cm & $^{34}$Si & $^{208}$Pb & 24.90                  &    23.15 & \cite{ogl00} \\
\hline
\hline\end{tabular}
\caption{\label{halflivestab} Half-lives for cluster emission from the 
light actinides calculated in the HFB mean-field approximation compared 
with the experimental data where available. Experimental data in each 
section correspond to one or two emitted nuclei as indicated.
The theoretical results include the ZPE correction, for the corresponding
values without it, see Fig. \ref{halflives}. In general terms the half-lives
without ZPE are one to two orders of magnitude larger than the ones computed
with the ZPE.
}

\end{table}

The previous section contains a complete analysis of two relevant 
examples of cluster decay in the actinides. Despite the differences 
in masses of the two isotopes considered there, it is clear that the 
same mechanism is responsible for CR in all the nuclei in this 
region. In this section, the results for the other isotopes are 
presented. We  concentrate on the CR fission barriers and on the 
half-lives for this very mass asymmetric decay. A thorough 
comparison of our theoretical  results with experimental data will 
be also discussed in this section.
 
In Fig. \ref{allbar} the fission barriers are presented as a 
function of the octupole moment $Q_3$ for all actinide nuclei in 
which cluster emission was  experimentally investigated. The height 
of the barrier remains roughly constant for all the considered 
nuclei and it only varies between 22 and 26 MeV. The saddle point is 
located at small octupole deformations $Q_3=20 \; \mathrm{b} ^{3/2}$ 
for light nuclei and it gradually shifts to the $Q_3=55 \;\mathrm{b} 
^{3/2}$ for the heaviest ones.

In some nuclei we have plotted curves for two or three possible mass 
distributions of the clusters obtained with Eq. (\ref{coulomb}). We 
do so also when the light fragment corresponding to the minimum of 
the CR valley does not agree with the mass of the cluster observed 
experimentally. In these cases we consider both clusters measured 
experimentally  and predicted by the model as possible decay 
products and we determine the corresponding CR barriers and 
spontaneous emission half-lives for all possible clusters. In some 
nuclei the mass of the emitted cluster has not been precisely 
identified by the experiment (e.g. for $^{236}$U). In these isotopes 
we have calculated barriers for all possible clusters as well. It is 
worth recalling that all the curves converge close to the saddle 
point. This is because at the saddle point a  difference of two 
neutrons or protons in any of the fragments do not change 
substantially the shape of density distribution or energy of the 
whole system. Therefore, the nucleus at the scission point can chose 
any of the decay channels. 

The flat shape of the hyperbolic curves of Eq. (\ref{coulomb}) at 
high deformation leads to important differences in the length of the 
barrier for each decay products. This is the reason for the few 
orders of magnitude differences in the half-lives of each of the 
cluster emission reactions.

Half-lives calculated for all considered cluster emissions are 
plotted in Fig. \ref{halflives}. Corresponding data are also 
presented in Table \ref{halflivestab}.  The theoretical results are 
compared with known experimental data where available. As can be 
seen in this Figure, measured half-lives are reproduced with an 
accuracy of two or three orders of magnitude in the most of the 
cases. This is the typical kind of agreement observed in all the 
other models of cluster emission. 

In Fig. \ref{halflives} we have presented half-lives calculated 
with the energy of the ``fusion" path 
given in Eq. (\ref{coulomb}). This classical expression does not 
include pure quantal effect of the ZPE. Therefore we have shown also the 
results with the ZPE calculated in the microscopic way subtracted 
from the Coulomb energy. Table \ref{halflivestab} contains only 
half-lives calculated it the latter manner.

In $^{226}$Ra we have found two parallel fission path after the 
saddle point. Such configuration of the PES suggest two modes of 
hyper-asymmetric fission  with two different clusters that can be 
produced. Results of half-lives on both paths show much faster decay 
with emission of experimentally observed $^{14}$C than with $^{20}$O 
determined in the other ``fusion" path.  

In a few cases ($^{226}$Ra, $^{230-232}$Th, and $^{230}$U) the 
clusters predicted by the HFB calculations are not observed 
experimentally. A possible explanation is that the half-lives of the 
experimentally observed decays are shorter by a few orders of 
magnitude than the not observed decays predicted by our 
calculations. This can be seen, for instance, for the $^{230}$Th 
decay in  Fig. \ref{allbar} where the energy curve for the 
experimentally observed cluster $^{24}$Ne decreases faster than for 
$^{22}$O calculated in the ``fusion" path. This produces a  decrease 
of the width and the height of the barrier and increases the 
tunneling probability.

\section{Conclusions}\label{sec5}

Cluster radioactivity is a very exotic kind of nuclear decay. It 
represents the bridge connecting the standard spontaneous nuclear 
fission phenomenon with $\alpha$ particle emission. We have shown 
here that this process can be understood as a kind of 
hyper-asymmetric spontaneous fission where the dynamics is governed 
by the shape of the potential energy surface given as a function of 
the quadrupole and octupole moments and computed in a fully 
microscopic way. The valley on the potential energy surface going 
through reflection asymmetric shapes has been identified as the 
responsible for the decay branch with emission of light cluster. The 
very high barriers involved in this process, reaching 25 MeV, result 
on extremely long half-lives for this decay. Scission point of the 
hyper-asymmetric fission is localized in the region of the saddle. 
The cluster is formed in the process of deforming a nucleus to a 
molecular shape with a heavy fragment close to the $^{208}$Pb 
nucleus. This approach is successful in predicting the mass and 
charge of the emitted particles as well as  half-lives in each of 
the isotopes where the process is experimentally known.

The standard method for calculating the potential energy surface in 
the fully microscopic mean-field model and determining the fission 
paths has been applied in this paper. The innovation of our 
description as compared to standard fission is the use of the 
octupole moment as the leading coordinate driving the system from 
its ground state configuration to hyper-asymmetric fission. This 
choice allows to determine easily fission path at the bottom of the 
valley and can be used in calculating half-lives.

A detailed analysis of the scission point has been performed. We 
have found the  continuous path for the transition from compound 
nucleus to two separated fragments. Applying constraints on the neck 
parameter is crucial in describing the potential energy surface in 
the region around scission point without loosing continuity. It has 
been proven that the whole up-going part of the fission path 
contribute to the energy barrier. At its end a neck is ruptured 
without change of the quadrupole or octupole moment, but with rapid 
decrease of energy. The rest of the barrier describing the energy of 
the separated fragments can be approximated by the classical 
expression for the Coulomb energy between two charged spheres.

\begin{acknowledgments} 

The work of LMR was supported by Ministerio de Ciencia e Innovaci\'
on (Spain) Grants Nos. FPA2009-08958, and FIS2009-07277, as well as 
by Consolider-Ingenio 2010 Programs CPAN CSD2007-00042 and MULTIDARK 
CSD2009-00064. The work of MW was supported by Ministerstwo Nauki i 
Szkolnictwa Wy\.{z}szego (Poland) under Grant No.\ N~N202~231137. 

\end{acknowledgments}

\appendix
\section{Convergence of the energy with basis size}
\label{appa}

\begin{figure}
\includegraphics[clip,width=0.7\columnwidth, angle=270]{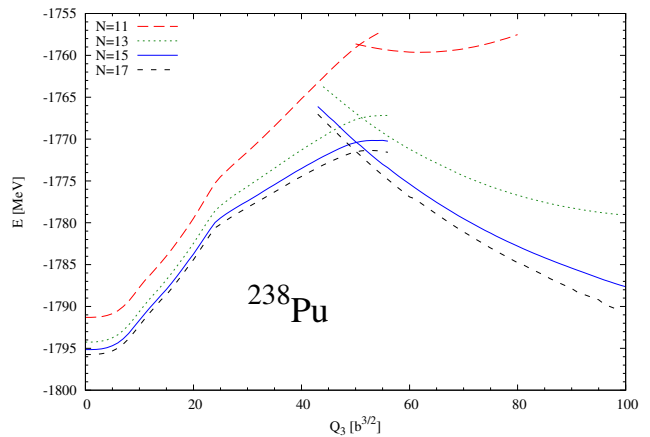}
\caption{\label{conv} 
Convergence of the computed energy with the basis size characterized 
by the basis size parameter $N$ (see text for details).}
\end{figure}

The axially symmetric harmonic oscillator (HO) basis used to expand 
the HFB quasiparticle operators is characterized by the number of 
shells chosen for the $z$ and perpendicular directions and by the 
corresponding oscillator lengths $b_{z}$ and $b_{\bot}$. The number 
of shells is restricted by the condition 
$\frac{1}{q}n_{z}+(2n_{\bot}+|m|)\leq N$, where $q$ and $N$ are 
parameters characterizing the basis. Along the perpendicular 
direction we take $N$ shells, (i.e. $2n_{\perp}+|m|=0,\ldots,N$) and 
along the $z$ direction we include up to $qN$ shells depending on 
the value of $2n_{\perp}+|m|$. In our calculations we have taken 
$q=1.5$ which is a good choice for elongated matter distributions 
extending along the $z$ direction and for $N$ (which determines 
roughly the number of shells present in the basis) we have taken the 
values 11, 13, 15 and 17 in order to check the convergence of the 
calculations with the basis size.

In all the cases the oscillator lengths $b_{z}$ and $b_{\bot}$ 
characterizing the basis are optimized as to minimize the energy for 
each value of the considered octupole moment. The results of the 
check for the nucleus $^{238}$Pu along the CR path are summarized in 
Fig. \ref{conv}. In this figure we have plotted the HFB energy as a 
function of $Q_{3}$ for all the values of $N$ used. We can see that 
as $N$ increases, the basis size increases and the energy (a 
variational quantity) decreases. The change in energy in going from 
$N$ to $N+2$ (as seen in the picture) decreases with increasing $N$. 
This change grows up with the quadrupole moment suggesting that very 
elongated shapes require bigger basis for a proper description. We 
clearly observe a nice converge of the energy as $N$ increase giving 
us confidence that the $N=17$ results are already a faithful image 
of results to be obtained with an infinite basis. However, this 
basis is still rather demanding of long computational time (CPU time 
is usually multiplied by a factor of eight when $N$ is increased by 
two)  for the extensive calculations considered in this paper and we 
have taken $N=15$ in all the calculations considered in the paper. 
This choice provides energies which are roughly parallel to the 
$N=17$ for octupole moments up to $Q_3=80$ b$^{3/2}$. Difference 
between $N=15$ and $N=17$ changes only from 0.6 MeV for $Q_3=0$ b$^{3/2}$ 
to 1.6 MeV for $Q_3=60$ b$^{3/2}$ and 2.0 MeV for $Q_3=80$ 
b$^{3/2}$. Thou, the results depending upon energy differences will 
not be affected by our choice. However, for octupole moments greater 
than $Q_3=80$ b$^{3/2}$ the $N=15$ and $N=17$ bases are no longer 
parallel (difference between energies for $Q_3=80$ b$^{3/2}$ and 
$Q_3=100$ b$^{3/2}$ grows by 1 MeV) suggesting bad convergence of 
the $N=15$ basis. Tests for bases beyond $N=17$ are hard to perform 
as very large bases show numerical instabilities in the evaluation 
of matrix elements of the interaction due to the finite accuracy of 
computer's floating point arithmetic. As discussed in the body of 
the paper, this region corresponds to two separate fragments where 
only the Coulomb repulsion energy changes with $Q_3$ and in this 
case it is better to use the classical expression for the Coulomb 
energy. 

Despite the too--high energy of the cluster radioactivity valley for 
N=13 its shape is already well defined. Both ``fission" and 
``fusion" paths have shapes that match exactly the ones of the 
larger bases. The cluster radioactivity phenomenon can be 
qualitatively  explained in this case although some quantities 
(barrier heights, half-lives) will be overestimated.

\section{Various Gogny forces}
\label{appb}
\begin{figure}
\includegraphics[clip,width=0.7\columnwidth, angle=270]{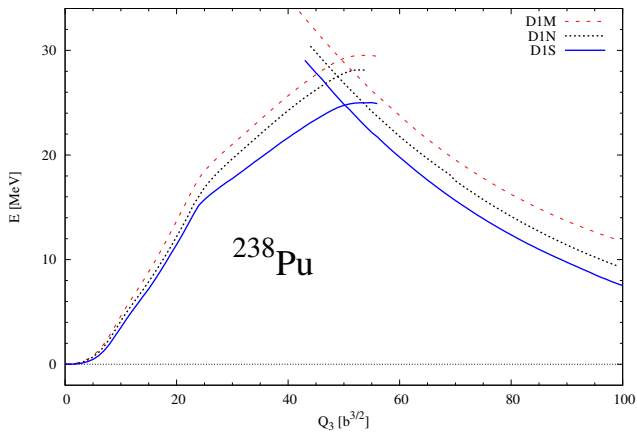}
\caption{\label{D1SNM} The hyper-asymmetric barrier in $^{238}$Pu calculated with different parametrizations
of the Gogny force: D1S, D1N, and D1M. }
\end{figure}

Recently, two new parameterizations of the Gogny force have been 
presented. They are called  D1N  \cite{cha08} and D1M \cite{gor09}. 
The parameters of D1N were constrained to reproduce, in addition to 
the standard requirements, the shape of a  realistic symmetric and 
neutron matter equations of state. The idea was to improve the 
properties of the force for neutron rich nuclei. The D1M 
parametrization also included the requirements of D1N but the 
parameters were fitted to the binding energies of the whole 
even-even nuclide chart and therefore its binding energy rms is 
outstanding. The properties of D1M have been tested in several 
scenarios \cite{D1Mtest,war10} with very promising results. 

It is interesting to test the performance of the three 
parameterizations in the very demanding scenario of CR and to this 
end we have plotted in Fig. \ref{D1SNM} the results for the PES of 
CR for the nucleus $^{238}$Pu obtained with the three 
parameterizations. The shape of the PES is similar in the three 
cases although the hyper-asymmetric fission barrier is 2 MeV higher 
in D1N and 3 MeV in D1M than in D1S. The predicted cluster is the 
same $^{30}$Mg nucleus in all three parameterizations. The changes 
of the barrier heights lead to an increase of the CR half-lives. For 
the D1M parameter set in the case of emission of $^{30}$Mg from 
$^{238}$Pu we get $\log(t_{1/2} [s])=34.33$  (without ZPE correction 
added) and for the D1N force $\log(t_{1/2} [s])=32.28$ in comparison 
with $\log(t_{1/2} [s])=30.42$ for the D1S parameterization.

The new Gogny parameterizations D1M and D1N provide a similar picture of 
the CR phenomenon although numerical calculations give slightly 
larger values of barrier heights and half-lives.




\bibliographystyle{model1a-num-names}

\end{document}